\documentclass[sigconf]{acmart}

\AtBeginDocument{%
  \providecommand\BibTeX{{%
    \normalfont B\kern-0.5em{\scshape i\kern-0.25em b}\kern-0.8em\TeX}}}

\copyrightyear{2026}
\acmYear{2026}
\setcopyright{cc}
\setcctype{by}
\acmConference[CHI '26]{Proceedings of the 2026 CHI Conference on Human Factors in Computing Systems}{April 13--17, 2026}{Barcelona, Spain}
\acmBooktitle{Proceedings of the 2026 CHI Conference on Human Factors in Computing Systems (CHI '26), April 13--17, 2026, Barcelona, Spain}
\acmPrice{}
\acmDOI{10.1145/3772318.3791637}
\acmISBN{979-8-4007-2278-3/2026/04}

\usepackage{color}
\usepackage{soul}
\usepackage{pifont}        
\usepackage{graphicx}
\usepackage{multirow}      
\usepackage{caption}
\usepackage{subcaption}

\newcommand\ie{\textit{i.e.},~}
\newcommand\eg{\textit{e.g.},~}
\newcommand\vs{\textit{vs.}~}

\newcommand\etal{\textit{et al.}~}

\newcommand\done[1]{}

\newcommand\changed[1]{#1}

\newcommand\p[1]{\paragraph{\textbf{#1}}}

\newcommand\degree{$^{\circ}$~}

\begin{document}

\title[PeriphAR: Fast and Accurate Real-World Object Selection with Peripheral AR Displays]{\textit{PeriphAR}: Fast and Accurate Real-World Object Selection with \\ Peripheral Augmented Reality Displays}

\author{Yutong Ren}
\affiliation{
  \institution{University of Michigan}
  \city{Ann Arbor}
  \state{Michigan}
  \country{USA}
}
\email{rentony@umich.edu}

\author{Arnav Reddy}
\affiliation{
  \institution{University of Michigan}
  \city{Ann Arbor}
  \state{Michigan}
  \country{USA}
}
\email{arnavr@umich.edu}

\author{Michael Nebeling}
\affiliation{
  \institution{University of Michigan}
  \city{Ann Arbor}
  \state{Michigan}
  \country{USA}
}
\email{nebeling@umich.edu}

\renewcommand{\shortauthors}{Y. Ren, A. Reddy, M. Nebeling}

\begin{abstract}
\changed{Gaze-based selection in XR requires visual confirmation due to eye-tracking limitations and target ambiguity in 3D contexts. Current designs for wide-FOV displays use world-locked, central overlays, which is not conducive to always-on AR glasses. This paper introduces \textsc{PeriphAR} \textit{(/peh-ree-faar/)}, a visualization technique that leverages \textit{peripheral vision} for feedback during gaze-based selection on a monocular AR display. In a first user study, we isolated text, color, and shape properties of target objects to compare peripheral selection cues. Peripheral vision was more sensitive to color than shape, but this sensitivity rapidly declined at lower contrast. To preserve preattentive processing of color, we developed two strategies to enhance color in users' peripheral vision. In a second user study, our strategy that maximized contrast of the target to the neighboring object with the \textit{most similar color} was subjectively preferred. As proof of concept, we implemented \textsc{PeriphAR} in an end-to-end system to test performance with real-world object detection.}
\end{abstract}




\maketitle


\section{Introduction}

\changed{Future AR glasses will require strategic compromises in display and sensing functionality to achieve optimal form factors conducive to always-on, pervasive AR \cite{Grubert2017}.
For example, the new Meta Ray-Ban Display smart glasses set the emphasis on wearable \textit{AI}, rather than \textit{AR}, featuring only a limited, monocular display compared to AR headsets such as HoloLens. While these smart glasses come with advanced input channels using ultra-wide cameras, directional microphones, and EMG wristbands, they essentially function as a heads-up display.
The lack of world-locked rendering makes real-world object selection, a fundamental interaction step in any AR experience \cite{bowman2021}, particularly challenging.}
\changed{A related problem with a near-eye display that projects an image slightly off to the side of one eye, thus not directly in the user's line of sight, is that it essentially creates a floating window in the user's periphery. Interacting with such a display requires frequent gaze shifts and focal length adjustments that can be distracting from real-world interaction \cite{Pfeuffer-CG21} as well as increasing digital eye strain \cite{Hirzle-CHI19,Hirzle-ETRA20}.}

\changed{In this paper, we introduce \textsc{PeriphAR} \textit{(/peh-ree-faar/)}, a \changed{visualization technique that leverages \textit{peripheral vision} for feedback during gaze-based selection on monocular AR displays}.
\changed{Being well established in XR} \cite{Feit-CHI17}, gaze-based selection was previously shown to naturally support intuitive hands-free interaction \cite{pai2019handsfree, pfeuffer2017gazepinch, sidenmark2020outline, sidenmark2019eyehead}. However, existing techniques typically rely on visual cues, such as eye cursors, gaze rays, or object highlights \cite{pfeuffer2017gazepinch, chen2023gazeraycursor}, to be directly overlaid in the user's central (or foveal) vision. Although prior work has shown that near-peripheral vision can present secondary information without disrupting tasks that require foveal attention \cite{janaka2022paracentral}, it has not been previously studied as a feedback channel for object selection. With peripheral vision being less sensitive to detail \cite{luyten2016hidden}, the main challenge is deriving a visual representation from real-world targets that encodes the most salient visual properties of the target so that it can be clearly perceived in peripheral vision.}

\changed{With \textsc{PeriphAR}, we make two contributions:
\textit{(1)} we present controlled experiments that provide empirical insights into effective peripheral cues for real-world object selection that enhance selection accuracy, reduce cognitive load, and increase user confidence;
\textit{(2)} we detail the design of a pipeline implementing two strategies to enhance color for peripheral vision, demonstrating that our strategy that maximized contrast of a target to the neighboring object with the \textit{most similar color} was preferred and yielded higher confidence.}

\section{Background and Related Work}

This research establishes principles of limited, peripherally glanceable AR displays that allow users to keep their visual attention in the real world and simultaneously perceive content in the display without having to focus on it.

\p{Glanceable AR Displays} There is an existing concept of glanceable AR presented by Lu \etal~\cite{Lu-VR20,Lu-VR21}.
In \cite{Lu-VR21}, they studied benefits and drawbacks of glanceable AR applications for different use cases at work, home, or on-the-go based on video and functional prototypes.
In \cite{Lu-VR20}, they evaluate interaction techniques designed to bring AR content from the peripheral to central vision. 
Finally, in \cite{Lu-CHI22}, they studied adaptable and adaptive AR content transition mechanisms for moving interface widgets to different locations in the user's physical context.
Their concept is related but different from ours. One important difference is that these studies assume more capable AR displays (modeled after HoloLens and Magic Leap, \ie with binocular views, world-locked rendering, and wide FOV). Importantly, they distinguish the periphery relative to the eye and to the head, and reserve the peripheral vision to anchor virtual objects with physical locations to present secondary information in collapsed views that will expand as the user brings them into their central vision through either head or eye movements.
In contrast, we aim to avoid the need for foveating towards the display, instead providing visual, peripheral cues in the monocular display so that users can perceive them while keeping their attention real-world.

Prior work has investigated lightweight, glanceable designs for AR glasses, highlighting challenges of limited fidelity and brightness \cite{rauschnabel2015ar,koulieris2019near}. Researchers proposed context-aware and adaptive interfaces to improve glanceability \cite{davari2022validating,Lindlbauer-UIST19}. One particularly promising direction explored dynamic color enhancement to maximize perceptual clarity under resource constraints. For example, Chroma tailored AR palettes for color-blind users \cite{tanuwidjaja2014chroma}, Chen \etal~\cite{chen2023imperceptible} applied imperceptible modulation for power saving, and Duinkharjav \etal~\cite{duinkharjav2022color} optimized rendering with perception-guided recoloring. These works illustrate how adaptive color feedback can reduce clutter or energy use. Our work builds on this literature but applies color enhancement specifically to peripheral cues to improve selection confirmation.

\p{Designing for Peripheral Vision}
Most existing designs for head-worn AR target central and foveal vision, largely neglecting peripheral vision.
Yet, peripheral vision excels at detecting motion and high-contrast changes, providing users with ambient awareness without requiring direct gaze shifts. Strasburger \etal~\cite{strasburger2011peripheral} characterize peripheral vision to be of lower spatial resolution but with a higher sensitivity to contrast, making it ideal for subtle but salient notifications. Building on this, Sun and Varshney~\cite{sun2018investigating} quantified the perception time in the far peripheral field for VR/AR, showing that users can register peripheral stimuli within milliseconds, even under complex visual loads.

The HCI literature contains few studies on peripheral vision, which can be classified as building custom hardware to create new types of displays \cite{Costanza-MobileHCI06,Luyten-CHI16}, studying display placement \cite{Arora-2025,HaynesS-IMWUT17}, and the level of detail of the information shown \cite{DiVerdi-ISMAR04,Gruenefeld-MobileHCI18,Gruenefeld-ISMAR18}.
Of particular interest to this research is the work by Luyten \etal~\cite{Luyten-CHI16} who created a display setup involving two peripheral displays mounted within the 20-degree extremes of the periphery to study the usable peripheral display area in terms of position and size relative to the eye, and the impact of shape, orientation, color, and animation on users' perception. They recommended \textit{(1)} using a limited set of simple, prominent shapes (\eg rectangles and circles in 2D), \textit{(2)} using primary, strongly contrasting colors (their participants correctly perceived red, green, and blue, but had challenges with yellow, purple, and orange), and \textit{(3)} using animation intentionally to communicate more complex information (\eg motion in the direction an arrow is pointing).

A new stream of research has leveraged peripheral channels for subtle, multisensory cues.
Ku \etal~\cite{ku2019peritext} used rapid serial visual presentation (RSVP) to deliver textual overlays that are readable using peripheral vision.
Trepkowski \etal~\cite{trepkowski2021multisensory} combined visual, auditory, and haptic signals to improve awareness under a narrow field-of-view (FOV). Parmar \& Silpasuwanchai \cite{parmar2023impact} studied attentional tunneling while walking, Janaka~\etal~\cite{janaka2022paracentral} explored paracentral visualization strategies, and Syiem \etal~\cite{syiem2021impact} showed reduced cognitive load during multitasking. This stream suggests peripheral vision is an effective communication channel; however its use to interact with AR content remains underexplored.

\p{Feedforward and Feedback in Visual Guidance}
\changed{Our framing of \textsc{PeriphAR} as a visual \emph{feedback} mechanism builds on prior work that distinguishes feedforward from feedback in interaction design. Vermeulen \etal~\cite{Vermeulen-CHI13} clarify feedforward as information presented \emph{before} an action is carried out to reduce Norman's Gulf of Execution, complementing but differing from feedback about the outcomes of completed actions. In pointing interfaces, Guillon \etal~\cite{Guillon-CHI15} introduce a design space of visual feedforward for target expansion, emphasizing that users must see how targets will expand \emph{prior} to movement to benefit from the technique. Sadasivan \etal~\cite{Sadasivan-CHI05} similarly use recorded eye movements as feedforward training cues that preview how experts would inspect complex aircraft structures, again focusing on anticipatory guidance rather than on-line correction.}

\changed{Recent XR work further sharpens this distinction. Yu \etal~\cite{Yu-CHI24} propose a design space for visual feedforward and \emph{corrective feedback} in XR motion guidance, defining feedforward as cues that specify desired future motion and feedback as cues that respond to a user's ongoing performance and highlight errors. In parallel, research on peripheral visual feedback shows that unobtrusive cues in the visual periphery can efficiently signal system state and guide attention without disrupting foveal tasks~\cite{Nikolic-HF01}. AR studies have leveraged such feedback to support continuous control and search: Wang \etal~\cite{Wang-JCISE24} compare central and peripheral visual feedback (and other modalities) for trajectory tracking in AR, and Richards \etal~\cite{Richards-SUI19} combine peripheral visual and vibrotactile feedback to help users locate dynamic virtual entities in augmented reality. Gaze-based targeting work likewise uses visual highlights and cursor changes as feedback about the system's current interpretation of the user's gaze selection~\cite{Fernandes-IJHCI25}. Following Yu \etal's distinction between feedforward and \emph{corrective feedback} in XR motion guidance~\cite{Yu-CHI24}, these cues are presented in response to the user's or system's \emph{current} state, rather than predicting future actions.}

\changed{We explicitly position \textsc{PeriphAR} as providing \emph{peripheral feedback} instead of feedforward: our peripheral proxies appear only after the system has interpreted a fixation on a real-world object, and they visualize “what the system currently thinks you are selecting'' in the monocular display. Users can then confirm or reject this interpretation without foveating toward the display.}

\p{Preattentiveness to Color and the Link to Peripheral Vision}

\changed{Classic work on preattentive vision shows that color is a particularly powerful feature for guiding visual search. In dense arrays, color singletons can be detected in parallel and ``pop out'' even when observers are not deliberately attending to them~\cite{nothdurft1993preattentive}. Later work refined this view by demonstrating that individual hues differ in how strongly they guide attention. For example, Andersen \etal found that red tends to be detected the fastest, followed by green, blue, and yellow, whereas orange and purple are less effective at improving search performance~\cite{andersen2019attentional}. Beyond low-level contrast, preattentive responses are also shaped by color categories themselves. Using peripheral oddball paradigms and visual mismatch negativity, Clifford \etal showed that across-category differences (e.g., blue \vs green) elicit stronger preattentive responses than equally large within-category differences~\cite{clifford2010color}. These results suggest that ergonomic design guidelines for color should consider not only contrast and luminance, but also the choice and categorical perception of specific hues.}

\changed{Preattentive color processing is closely linked to peripheral vision. Visual search theories argue that feature maps for color, orientation, and motion are computed across the visual field, providing a coarse salience landscape that can guide eye movements toward promising locations~\cite{wolfe2018visual}. Empirical studies show that color signals support rapid detection even when targets are presented outside the fovea. For example, Khomeriki and Lomashvili reported that under peripheral viewing, red stimuli are detected faster than green or blue, and that this advantage interacts with task complexity and spatial location in the visual field~\cite{khomeriki2025attention}. Together with classic findings on preattentive filling-in and texture segmentation~\cite{nothdurft1993preattentive}, this work indicates that color-based guidance is available in peripheral vision, albeit at lower spatial resolution than in central vision.}

\changed{Our work builds on this literature by exploiting peripheral color sensitivity in an AR context. \textsc{PeriphAR} presents proxy objects in the user's peripheral rather than foveal vision, and our studies suggest a predisposition to color and similar ordering of color effectiveness (with red targets noticed fastest, and green/yellow slower), consistent with preattentive color findings in both central and peripheral vision~\cite{andersen2019attentional,clifford2010color,khomeriki2025attention,nothdurft1993preattentive}. We leverage these insights when reasoning about how color-enhanced peripheral proxies can support rapid confirmation of gaze-based selections on low-FOV AR glasses, and we return to them in our discussion of future extensions for multi-colored, real-world targets (Sec.~\ref{sec:discussion}).}

\section{Peripheral AR Display Simulation}
\label{sec:simulation}

\begin{figure*}[ht]
  \centering
  \begin{minipage}[t]{0.49\textwidth}
    \centering
    \includegraphics[width=\textwidth]{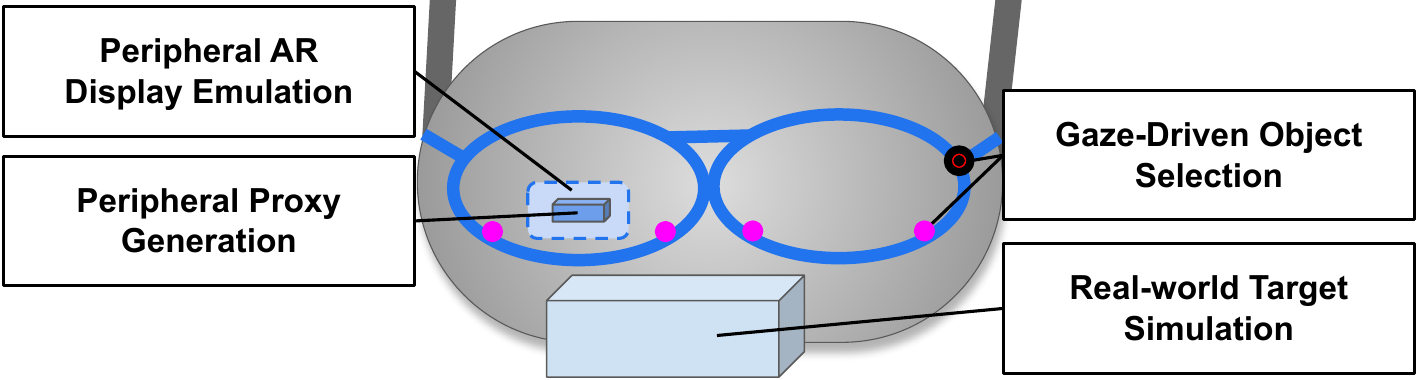}\vspace{-2ex}
    \caption{Our system consists of four components: \textit{(1) peripheral AR display emulation} simulates monocular AR glasses with limited FOV on Quest Pro, while \textit{(2) gaze-driven object selection} uses eye tracking and video passthrough for object selection; \textit{(3) real-world target simulation} enables simulation of target objects using virtual 3D shapes and models in world space, while \textit{(4) peripheral proxy generation} creates a visual representation of the real-world target on the simulated display screen. The system was used in two studies to determine key object properties for visually salient target representations.}
    \Description{Diagram of the PeriphAR system architecture. A glasses illustration represents simulated AR hardware. Four labeled components surround it: ``Peripheral AR Display Emulation'' and ``Peripheral Proxy Generation'' on the left, and ``Gaze-Driven Object Selection'' and ``Real-world Target Simulation'' on the right. Arrows connect each label to the glasses, showing how the components interact to support peripheral AR object selection.}
    \label{fig:system}
  \end{minipage}
  \hfill
  \begin{minipage}[t]{0.49\textwidth}
    \centering
    \includegraphics[width=\textwidth]{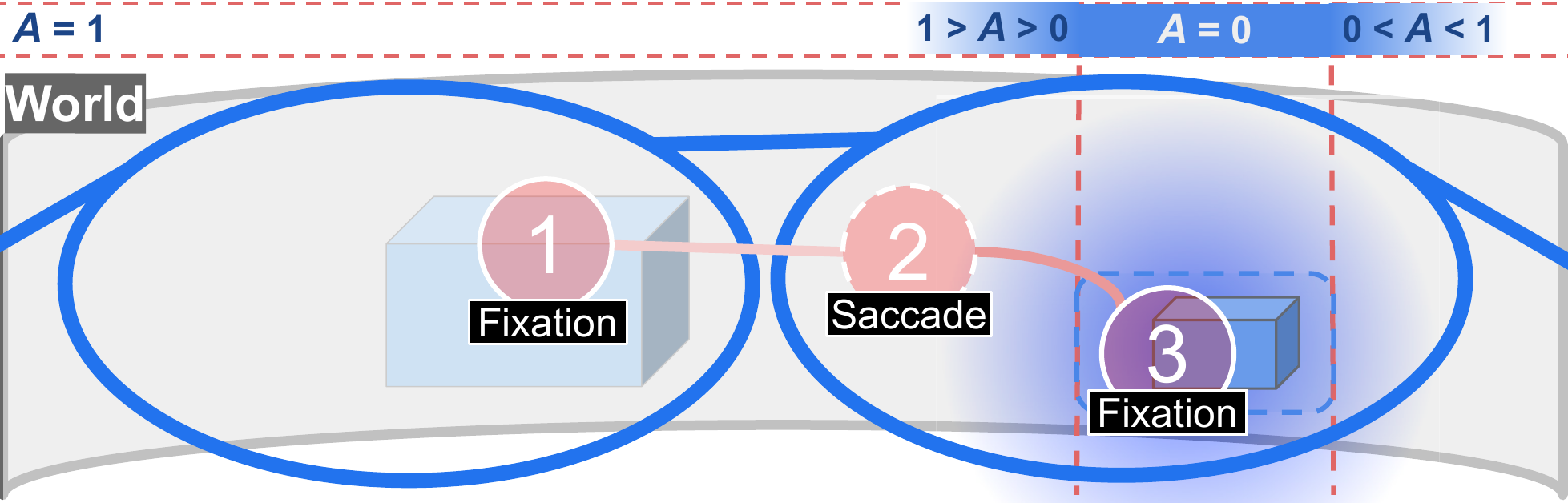}\vspace{-2ex}
    \caption{Our system measures \textit{display peripherality} based on gaze dynamics: a design that requires less time to look at, and fewer gaze transitions into, the display means it provides better peripherality. By structuring the visual field into a gradient (blue) around the simulated display, we compute the ambient value ($A$) of each gaze sample as: \textit{(1)} $A=0$, when the vector points \textit{in-world} and away from the display, \textit{(2)} $0<A<1$, when in the blue zone modulated within 10\degree in each direction around the display, or \textit{(3)} $A=1$, when \textit{in-display} and intersecting with the simulated display's screen area (dotted blue box).}
    \Description{Diagram illustrating how PeriphAR measures display peripherality from gaze dynamics. A wide horizontal scene labeled ``World'' is shown with a simulated rectangular display on the right. Three numbered circles mark gaze events: (1) Fixation on an object in the world to the left of the display, (2) Saccade as gaze moves toward the display, and (3) Fixation on the simulated AR display. Along the top, colored bands and labels show how the ambient value A is computed for each gaze sample: A=0 when gaze is in-world and away from the display, 0<A<1 in the blue transition zone within 10° around the display, and A=1 when gaze is in-display and intersects the display area (highlighted by a dotted blue rectangle).}
    \label{fig:ambient}
  \end{minipage}
  \vspace{-2ex}
\end{figure*}

\changed{To simulate a peripheral AR display and support the design of our \textsc{PeriphAR} feedback technique}, we developed a system based on Quest Pro, a current MR headset with built-in support for video passthrough and eye tracking. As illustrated in Fig.~\ref{fig:system}, the four main components of our system are: \textit{(1)} \textbf{peripheral AR display emulation} to gracefully degrade Quest Pro's display and rendering capabilities and simulate monocular AR glasses, \textit{(2)} \textbf{real-world object simulation} to generate virtual objects as targets and control their number, size, and distance in our experiments, \textit{(3)} \textbf{gaze-driven object selection} to enable object selection using eye tracking for both virtual objects generated by our simulation and physical targets detected by our system, and \textit{(4)} \textbf{peripheral proxy generation} to create a visual representation of a targeted virtual or real-world object as feedback in the simulated peripheral display.

\p{Display Emulation}
Simulating AR glasses is complex. For our experiments, we needed to control key parameters such as display position and size relative to a user's visual field on Quest Pro, but not accurately model all the optics (\eg luminance and brightness under different environmental conditions).
We scoped our display emulation to meet two requirements:
\textit{(i)} \textbf{control the simulated display's field of view and rendering capabilities} separately from the Quest Pro MR display with video passthrough rendering; and
\textit{(ii)} \textbf{emulate a monocular display context} that renders an image of the selected target in the simulated display only for the right eye.

Our system provides visual controls to adjust the position and size of the display area. We added means to simulate lower FPS and adjust the display image in terms of resolution and color. As part of the calibration in our experiments, we placed the display area in each user's periphery so that they could still see all four corners of the simulated display when foveating towards it. We consistently chose a 20\degree horizontal field of view and simulated the display at VGA resolution with RGB color at 30 FPS for most parts of the experiment (the text and shape conditions used grayscale). This yields a limited AR display that falls within the spectrum of current smart glasses. \changed{For example, the recently released Meta Ray-Ban Display glasses feature a 600x600 pixel, full-color RGB, monocular display in the right lens with a 20-degree field of view, and a 90Hz refresh rate (content refreshes at 30 FPS).}

\begin{figure*}[t]
  \centering
  \begin{minipage}[t]{0.55\textwidth}
    \centering
    \includegraphics[width=\linewidth]{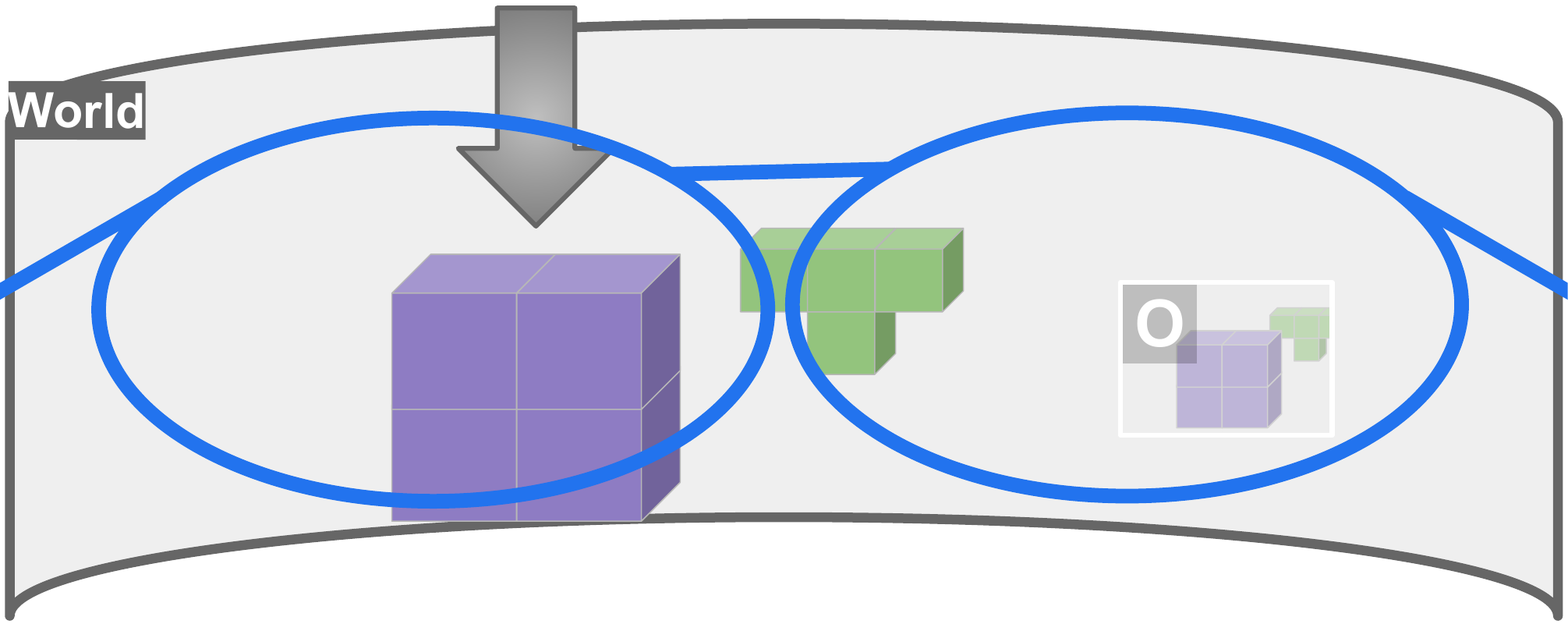}\vspace{-2ex}
    \caption{The tetris game in Study~1 simulated real-world targets using virtual objects in the shape of tetrominoes in four display conditions: the \textit{snapshot} condition combined the text label (O), square shape in color (blue O), and surrounding objects (green T). The \textit{text}, \textit{color}, and \textit{shape} conditions isolated these attributes as shown on the right.}
    \Description{Illustration of the Tetris-style task in Study 1. A wide “World” strip shows a large purple tetromino block on the left and smaller green tetromino blocks on the right. Two blue ovals overlay the scene: the left oval highlights the in-world target region around the purple block, and the right oval highlights the display region where a smaller, faded snapshot of the target and nearby blocks appears. The figure emphasizes that the game uses tetromino-shaped virtual objects as in-world targets and that different display conditions (snapshot, text, color, shape) present these attributes in the peripheral display.}
    \label{fig:tetris}  \end{minipage}
  \hfill
  \begin{minipage}[t]{0.43\textwidth}
    \includegraphics[width=\linewidth]{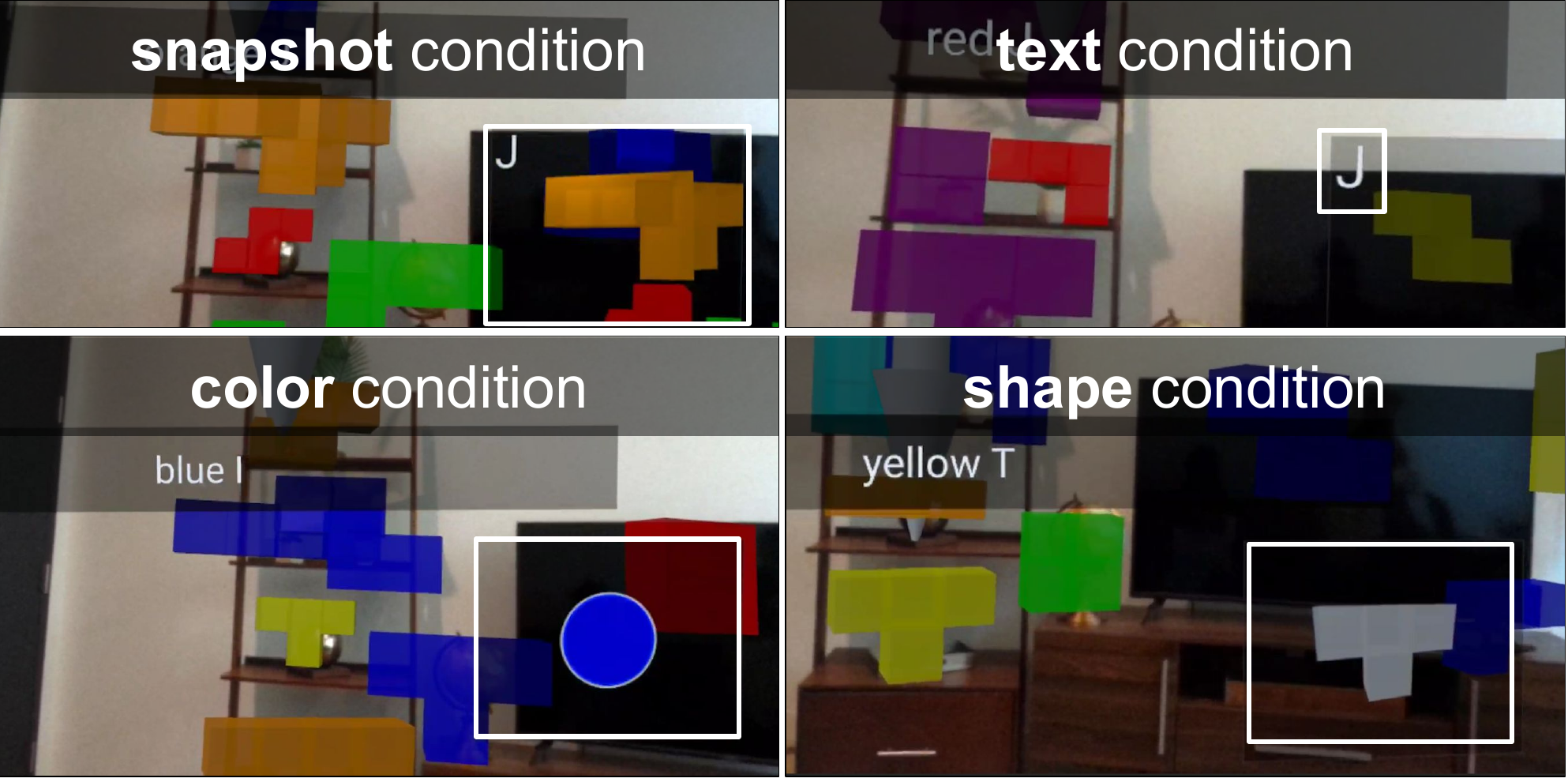}\vspace{-2ex}
    \caption{The four display conditions used in the tetris game: \textit{snapshot} with the text label and image of the target with surrounding shapes; \textit{text} with only the text label; \textit{color} with only a circle filled in the target's color; \textit{shape} with only the target's shape isolated in gray.}
    \Description{Four screenshots of the Tetris game illustrating different display conditions: (1) Snapshot condition shows the target tetromino highlighted with both a text label and a small image of the shape, along with surrounding blocks. (2) Text condition shows only a text label identifying the target block. (3) Color condition shows a circle filled with the target's color next to the block. (4) Shape condition shows the target's shape in gray outline.}
    \label{fig:tetris2}
  \end{minipage}
  \vspace{-2ex}
\end{figure*}

\p{Target Simulation}
\changed{For our experiments, we wanted to be able to simulate real-world targets using virtual 3D shapes and models registered in world space, serving as stand-ins for real objects in the physical environment. For example, our first study used real-world targets in the form of tetromino blocks (Fig.~\ref{fig:tetris}) (with seven different colors and shapes), and our second study used 3D fruit models (with similarly looking colors and shapes). This made it possible to observe gaze dynamics with respect to virtual targets whose real-world appearance and in-display representation we could fully control. It also meant that we could run experiments without being impacted by object recognition and segmentation errors.}

\changed{To test our \textsc{PeriphAR} visualization technique on real objects in the physical environment, we later developed an end-to-end system using YOLO11n~\cite{khanam2024yolov11} object detection and segmentation (Sec.~\ref{sec:end-to-end}). This served as a proof of concept that that our pipeline could be extended with computer vision techniques and used outside the lab, but was not meant as a generalization from our controlled studies.}

\p{Gaze-Driven Selection}
\changed{Our system implements gaze classification via velocity threshold identification (I-VT) for saccades and dispersion threshold identification (I-DT) for fixations \cite{Salvucci-ETRA00}, with fixed thresholds established in prior work \cite{Sendhilnathan-UIST22}. For durations, we used a 30ms maximum for saccades and a 50ms minimum for fixations in Study~1. In Study~2, we used 45ms minimum fixation duration to compensate for the delay in peripheral proxy generation.}

\changed{Our system further classifies gaze samples to determine display peripherality as a metric for comparison and proxy for glanceability.
The key idea is that a display condition is better if it effectively reduces eye movements, since it would make it easier to process the visual feedback rendered in the peripheral display and match it to the real-world target \textit{without} foveating towards the display. Thus, we were interested in detecting where fixations and saccades happened relative to the display.}

\changed{As illustrated in Fig.~\ref{fig:ambient}, our system structures the user's visual field into a gradient around the display, leading to clear-cut \textit{real-world} \vs \textit{in-display} attention zones, and a \textit{world/display} transition zone in between. 
Our display peripherality metric, then, consists of the combined durations of fixations on objects in the world, as opposed to fixations on the display, as well as the frequency of world/display transitions measured in terms of saccades between real-world and in-display fixations.}
We define an AR display to be of higher peripherality the higher the time real-world ($tW$), the lower the time in-display ($tD$), and the shorter and the fewer the transitions. We note that, when comparing different display conditions, key performance indicators depend on the complexity of the content shown in the display. That is, when comparing more complex displays, we can also expect higher values for the in-display metrics, leading to lower real-world to in-display ratios, hence decreased glanceability.

To determine intersections of the gaze with real-world targets, our system implemented a version of the flashlight technique often used in 3D contexts \cite{Liang-CG94,Pfeuffer-CG21}.
The system would lock onto the closest intersecting target and render a visual representation of it in the simulated display. In addition to visual cues rendered in the simulated display, we also added audio cues by playing distinct sounds each time the target has changed and when selection was confirmed via controller button press.
\changed{We used a controller-based confirmation rather than hand tracking to circumvent hand tracking errors and allow participation of users who were less familiar with XR technology.}


\p{Peripheral Proxy Generation}
We developed techniques to generate visual representations of real-world objects targeted by the user and used them to create four display conditions in Study~1 and three display conditions in Study~2, detailed in the next sections. The studies had different, yet complementary goals.
In Study~1 (Sec.~\ref{sec:study1}), we wanted to isolate key object properties with respect to display peripherality based on controlled experiments with virtual objects inspired from the popular tetris game.
In Study~2 (Sec.~\ref{sec:study2}), we wanted to compare different strategies for generating the peripheral proxy and assess if it can be interpreted peripherally without gaze shifts to the simulated display.
Finally, we developed an end-to-end system as a proof of concept (Section~\ref{sec:end-to-end}), employing real-time object detection and peripheral proxy generation from Quest Pro's video passthrough, and produced results for realistic environments.

\section{Study~1: Peripheral Sensitivity}
\label{sec:study1}


The goal of our first study was to condition the AR display design in terms of visual features--text, color, and shape---and study how each individually contributes to display peripherality.
To maximize control and balance visual features, we simulated targets using virtual objects in a game we developed inspired by Tetris.
For the experiment, we wanted virtual objects that can approximate the shape of real-world objects while being able to control and measure the differences.
Cuboids are commonly used in studies that segment real-world scenes (\eg \cite{Mousavian-CVPR17}), which motivated us to develop the Tetris game (Figure~\ref{fig:tetris}). Although the chosen colors in the game are less representative of the appearance of real-world objects, the selection of tones and shades in the color spectrum led to smaller and larger perceptual differences, which we considered realistic.

\p{Task}
Each trial involved selecting one of ten tetrominoes which were generated by our system using a combination of seven shapes (I or `straight,' O or `square,' T, L, J, S, and Z) and seven colors (cyan, blue, purple, green, yellow, orange, red) common to Tetris. Generated tetrominoes were unique in that the same color \textit{or} shape were possible, but no more than one tetromino of any color was present in a task. 

All tetrominoes were placed in fixed physical locations in front of participants who remained seated for control.
The system randomly selected each tetromino exactly once and announced it verbally (e.g., ``red T''), marking it with an arrow. We used this as an attention guiding technique \cite{Renner-3DUI17} with the goal of minimizing search time and reducing user effort to targeting, similar to when someone knows what object they are looking for and roughly where it is in the environment.
We opted for a continuous task design in that the next tetromino was marked and announced immediately after the correct one was selected.
As recommended in \cite{Bergstrom-CHI21}, we considered a discrete task design, but pilots found it too disruptive to realistically capture world/display transitions and make task completion time a factor. It would also have decreased the performance aspect of the game, which challenged participants to be accurate, yet efficient, by selecting the correct tetrominoes quickly.

\p{Procedure}
Participants conducted three blocks of 10 trials per block under four conditions (randomized and counterbalanced):
\textit{(1) snapshot} showed a colored image of the target shape with surrounding shapes as well as a text label with the shape's letter; 
\textit{(2) text} only showed the text label without color or shape, 
\textit{(3) color} just showed a circle in the color of the target, and
\textit{(4) shape} only showed the target's shape without color (Figure~\ref{fig:tetris2}).
The \textit{snapshot} condition functioned as a baseline with all features; the others were control conditions that isolated the individual characteristics to see how much each feature in isolation contributed to user performance.

\p{Participants}
In total, 32 participants completed the study (17 female, 15 male; ages 18--65, $M{=}38.28$, $SD{=}12.04$).
Participants' AR/VR experience varied from no prior experience (8 participants) to ten years ($M{=}1.72$ years, $SD{=}2.51$ year), using AR/VR on average 0.79 days per week ($min{=}0$, $max{=}5$). Screening confirmed that the participants were at least 18 years old and had not previously experienced motion sickness when using VR. Clear vision was required, wearing contact lenses rather than glasses, if needed, not to interfere with the eye tracking of Quest Pro.
Each study session lasted between 60 and 90 minutes. 
Participants were compensated \$75 USD per hour.

\p{Calibration \& Pre-Study Questionnaires}
Each participant performed individual calibration of Quest Pro using suggested fit adjustments and IPD values, then went through the eye tracking calibration procedure to build an optimized gaze model for each user. After successfully testing the calibrated model, participants completed a pre-study questionnaire asking for their perception of the most common symptoms of motion sickness (SSQ) and digital eye strain (DES) per Hirzle \etal \cite{Hirzle-CHI21}.
After each condition, we administered an abbreviated questionnaire to monitor any symptoms. Throughout the study, we followed best practices and exercised caution to be sensitive to any issues noted by the participants. Although no participants experienced motion sickness, one participant felt mild symptoms of digital eye strain after calibration, but the symptoms disappeared soon after and they completed the study.
Before each condition, we calibrated our system to ensure that the simulated display was positioned correctly in the participants' visual field and the gaze cursor was tracked accurately.

\p{Data Collection \& Analysis}
We used our system to record user sessions and produce real-time analytics based on our peripherality metrics. The record/replay system allowed us to collect and analyze detailed logs of participants' eye movements per task, block, and condition. We securely stored participants' anonymous experiment data and developed scripts in JavaScript and R to pre-process data from successful trials and perform statistical analysis. In post-hoc analysis, we compared the display conditions in terms of logged statistics and subjective ratings collected in post-task questionnaires.

\begin{table*}[t]
\resizebox{\linewidth}{!}{%
\begin{tabular}{l|lll|l|l|lll|llll}
\textbf{condition} &
  \textbf{t {[}s{]}} &
  \textbf{tD {[}s{]}} &
  \textbf{tW:tD} &
  \textbf{transitions} &
  \textbf{errors} &
  \textbf{satisfied} &
  \textbf{efficient} &
  \textbf{confident} &
  \textbf{comfortable} &
  \textbf{at right time} &
  \textbf{felt supported} &
  \textbf{felt in the way} \\ \hline
\textbf{snapshot} &
  33.3 &
  4.9 &
  4.1:1 &
  10 &
  1.31 &
  5.69 &
  5.03 &
  5.44 &
  4.81 &
  5.47 &
  4.97 &
  3.38 \\
\textbf{text} &
  33.6 &
  5.4 &
  3.6:1 &
  11 &
  \textbf{1.06} &
  5.91 &
  5.56 &
  \textbf{5.91} &
  5.19 &
  5.41 &
  5.25 &
  2.66 \\
\textbf{color} &
  \textbf{30.4} &
  3.0 &
  \textbf{6.0:1} &
  \textbf{7} &
  1.31 &
  \textbf{6.06} &
  \textbf{5.69} &
  5.88 &
  \textbf{6.13} &
  \textbf{6.09} &
  \textbf{5.91} &
  \textbf{2.09} \\
\textbf{shape} &
  34.8 &
  5.8 &
  3.7:1 &
  11 &
  1.64 &
  5.97 &
  5.34 &
  5.63 &
  5.22 &
  5.31 &
  5.00 &
  3.13 \\
\end{tabular}%
}
\caption{Peripherality metrics and post-task ratings for our Tetris game (best results in bold): mean task completion times ($t$) and time spent \textit{in-display} ($tD$), time \textit{real-world} \vs \textit{in-display} ratio ($tW$:$tD$), average number of world/display transitions and errors, and 7-point Likert scale ratings of task performance (satisfied, efficient, confident) and display quality (comfortable, at the right time, felt supported, felt in the way).
The \textit{color} condition was the fastest on average and achieved the best \textit{real-world} \vs \textit{in-display} gaze ratio, fewest transitions, and the best ratings in terms of display comfort, timing, feeling supported, and not feeling in the way.}
\Description{Table comparing four display conditions--snapshot, text, color, and shape--in the Tetris game study. Metrics include task completion time, time spent in display, ratio of in-world to in-display gaze, number of transitions, and errors. It also includes 7-point Likert ratings of task performance (satisfaction, efficiency, confidence) and display qualities (comfort, timing, support, obstruction). The color condition shows the best overall performance with the fastest completion time, fewest transitions, and highest ratings for comfort, support, and low obstruction.}
\label{tab:tetris}
\vspace{-4ex}
\end{table*}

Table~\ref{tab:tetris} presents key findings in terms of task performance recorded by the system and subjective performance based on post-task questionnaires on preference for condition, perceived task performance and display peripherality.

\p{Objective Performance}

We discarded the first block in all conditions that was used as training.
For statistical analysis, we ran Mauchly's test of sphericity against condition, then one-way repeated-measures ANOVA (with Huyn-Feldt correction if the sphericity could not be assumed) and, if significant, post-hoc pairwise t-tests with Bonferroni correction. Significant results mean $p < 0.05$.
The condition had a significant effect on task completion time, with \textit{color} being significantly faster than the \textit{shape} condition.
Overall, the fastest conditions per participant were \textit{color} ($17\times$), \textit{snapshot} ($6\times$), \textit{shape} ($5\times$), and \textit{text} ($4\times$).
The \textit{color} condition also outperformed the other conditions on average spending the least amount of time foveating towards the display, with significant differences between \textit{color} and all other conditions.
The \textit{tW:tD} ratios were the best for the \textit{color} condition, followed by \textit{snapshot}, then \textit{shape}, and finally \textit{text}.
The \textit{color} condition had the fewest world/display transitions on average, with significant differences between \textit{color} \vs \textit{shape} and \textit{color} \vs \textit{text}.
It also had the best \textit{real-world} \vs \textit{in-display} gaze ratio and was rated the best in terms of display comfort, timing, feeling supported, and not being in the way. It also had the best ratings for mostly looking at the world \vs the display.

\begin{figure}[t]
    \centering
    \includegraphics[width=.85\linewidth]{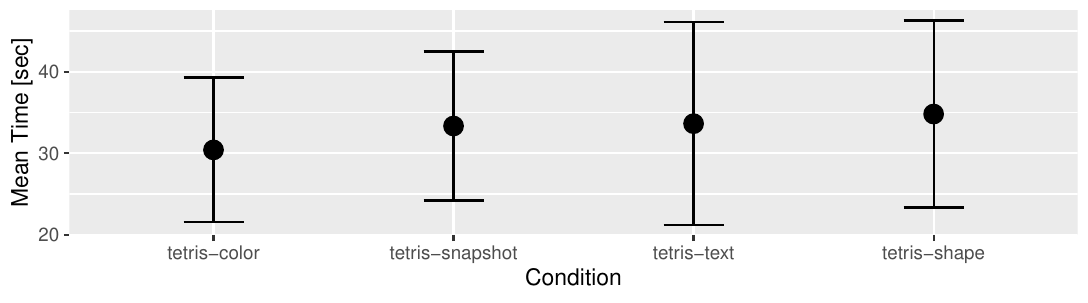}\vspace{-2ex}
    \caption{Mean task completion times in ascending order from left to right. Error bars show standard deviation.}
    \Description{Dot plot showing mean task completion times for four Tetris conditions: color, snapshot, text, and shape. Each condition is shown as a black dot with vertical error bars for standard deviation. The color condition has the lowest mean, while the other three are slightly higher and close together.}
    \label{fig:tetris_t}
    \vspace{-4ex}
\end{figure}

In terms of the total number of selection errors, the \textit{shape} condition had the highest number of errors for all participants ($total=105$, $mean=1.64$). The \textit{snapshot} and \textit{color} conditions performed similarly (both $total=84$, $mean=1.31$), while \textit{text} had the fewest errors ($total=68$, $mean=1.06$).
For further analysis, we replayed study sessions to analyze user behavior when errors occurred.
There was a fairly even spread of wrong color \vs shape selections for \textit{shape} ($23\times$ wrong color, $22\times$ wrong shape, $60\times$ both wrong), suggesting that shape alone may be insufficient. The \textit{color} condition more often led to selecting the wrong shape ($9\times$ wrong color, $33\times$ wrong shape, and $42\times$ both wrong). The \textit{snapshot} condition reversed the wrong color \vs shape ratio ($21\times$ wrong color, $11\times$ wrong shape, $51\times$ both wrong). The \textit{text} condition had the fewest errors ($16\times$ wrong color, $6\times$ wrong shape, $46\times$ both wrong).
The top-3 wrong colors were selecting orange instead of yellow ($10\times$), purple instead of teal ($6\times$), green instead of blue ($5\times$), and yellow instead of green ($5\times$). The top-3 wrong shapes were selecting J instead of L ($12\times$), L instead of J ($11\times$), and S instead of Z ($4\times$). Participants made the fewest errors with red (only once selecting teal instead) but confused all shapes at least once.

\p{Subjective Performance}

For statistical analysis, we used Friedman rank sum test and, if significant, post-hoc pairwise Wilcoxon signed rank test with Bonferroni correction.
Participants tended to rate their task performance the best for the \textit{color} and \textit{text} conditions, but the differences were not significant.
For the questions on the display, the \textit{color} condition was significantly rated the highest. We found significant differences between \textit{color} and all other conditions in terms of whether it was easy to understand the information shown on the display and whether it felt comfortable to look at the display.
It was also rated significantly better than \textit{snapshot} and \textit{shape} in terms of whether the information on the display was displayed at the right time and whether the display felt like it was in the way.
When asked whether the participants felt they mostly looked \textit{real-world} \vs \textit{in-display}, there was a trend towards \textit{real-world}, with the \textit{color} and \textit{shape} conditions rated the highest, but the differences were not significant.

\p{Reflection \& Takeaways}

From our experiment observations and participant feedback, two main themes emerged:
\textit{(1) Perceptual efficiency of color:}
Participants favored the color condition for its efficient communication of information. Particularly when perceived peripherally, color was the most potent feature and it provided a manageable amount of information, making it easier for participants to interpret.
\textit{(2) Limitations of color alone:}
Some participants encountered difficulties when nearby tetrominoes had similar colors, prompting reflections on the limitations of relying solely on color cues on a peripheral display.
Participants suggested combining color with other features such as shape. Some articulated benefits of the snapshot condition, as it showed more context, while others maintained that it was confusing.


\section{Peripheral Proxy Generation}
\label{sec:pipeline}

Given the findings from Study~1---that participants favored the color condition for its perceptual efficiency, but struggled when nearby objects had similar colors---we devised a technique that generates a peripheral proxy from a real-world object targeted by the user's eye gaze and makes differences to similar objects nearby more visible. Our technique is based on a color enhancement algorithm that shifts key color attributes--luminance, saturation, and chromaticity--to make the proxy more noticeable in peripheral vision while remaining consistent with the real-world object.

\begin{figure}[htbp]
  \centering
  \includegraphics[width=\linewidth]{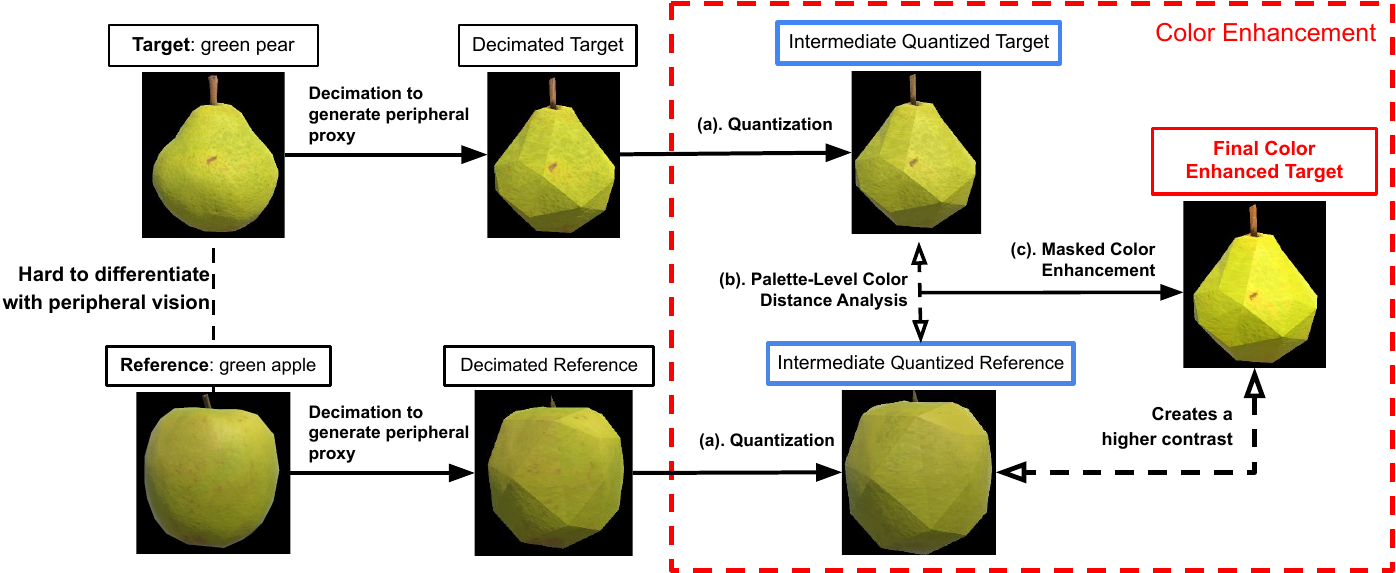}\vspace{-2ex}
  \caption{Color enhancement pipeline. 
    \textit{(a)} The target and reference undergo color quantization ($k{=}7$). 
    \textit{(b)} Palette‐level color distances are computed between the intermediate quantized target and reference. 
    \textit{(c)} The intermediate quantized target is color‐enhanced based on the computed palette‐level distances, producing the final texture for the peripheral proxy.}
  \Description{The figure shows a step-by-step pipeline comparing a green pear (target) and a green apple (reference). On the left, both fruits appear as photos and then as simplified blocky versions. In the center, these are further reduced into flat, polygon-like color proxies. Arrows connect them into a process where the pear and apple are compared. On the right, the pear is shown again but with brighter, more saturated colors, creating stronger contrast against the apple. The flow highlights how the target image changes from natural to simplified, then finally to a color-enhanced proxy.}
  \label{fig:color enhancement pipeline}
  \vspace{-3ex}
\end{figure}

\p{Color Enhancement Algorithm} As shown in Fig. ~\ref{fig:color enhancement pipeline}, our algorithm takes two inputs: a \textit{target} image for enhancement and a \textit{reference} image against which the target is contrasted.
The images are transformed through a pipeline in three steps: \textit{(1)} quantization, \textit{(2)} palette-level color distance analysis, and \textit{(3)} masked enhancement of the target. We detail the algorithm in Appendix \ref{app:algorithm} but give an intuitive explanation below.

Consider the green pear (target) and the green apple (reference) shown in Fig.~\ref{fig:color enhancement pipeline}. The green pear has a lighter and more yellowish green than the green apple. The differences are noticeable in foveal vision but much harder to detect in peripheral vision. The algorithm first simplifies the green pear's texture into a small set of dominant colors (quantization), then enhances lightness, saturation, and chromaticity to maximize contrast with the green apple (the quantization step and the full pipeline will be explained in detail below). The resulting green pear proxy appears brighter and more distinct, making it easier to recognize peripherally. If the target is darker than the reference, the algorithm performs only quantization and skips enhancement, since further darkening would reduce visibility in the periphery.



\p{Technical Pipeline}
\changed{Building on the gaze classification and display emulation described in Section~\ref{sec:simulation},} we implemented a pipeline that connects real-world fixation to peripheral proxy generation and subsequent rendering. Once a valid fixation is detected on a real-world object, our system designates it as the \emph{target} and extracts its texture as the target image from a decimated version of the object model. Decimation reduces geometric detail while retaining the outline shape, ensuring that proxies remain lightweight for peripheral rendering.
Next, a \textit{reference} image is obtained using one of two strategies described below. Both images are color enhanced by our algorithm, which adjusts the target's appearance relative to the reference to increase salience of the peripheral proxy while preserving visual consistency with the target.
The final color enhanced target is then displayed in the peripheral view, completing the feedback loop from gaze fixation to proxy confirmation.

\p{Strategy 1: Snapshot as Reference}
The first strategy assumes that users rely on the surrounding context of the real-world object. When a valid fixation is detected, \textsc{PeriphAR} registers the real-world object as the target and extracts a localized screenshot centered on it, including adjacent objects and background, similar to the \textit{snapshot} condition in Study~1. The screenshot serves as the reference image. By analyzing the screenshot's color distribution, the algorithm adjusts the target's luminance, saturation, and chromaticity to increase contrast with its surrounding objects captured in the screenshot. We refer to this as the \textit{screenshot} strategy.

\p{Strategy 2: Most Similar Object as Reference}
The second strategy goes beyond the first by identifying the one nearby object that is most likely to be confused with the target. Upon fixation, \textsc{PeriphAR} compares the target's color to its neighbors in \textsf{CIELAB} space using the \textsf{CIEDE2000} color-difference formula\changed{~\cite{sharma2005ciede2000}}, selecting as reference the adjacent object with the smallest color distance---i.e., the object most similar in color and therefore most likely to be confused with the target. The algorithm then enhances the target relative to this reference, adjusting luminance, saturation, and chromaticity to maximize perceptual contrast to this reference. We refer to this as the \textit{most similar color} (MSC) strategy.\\

\changed{\textit{Screenshot} optimizes contrast against the average context, whereas \textit{MSC} optimizes contrast against the worst-case distractor. 
In our second study, we tested both strategies (average \vs worst-case) to generate a peripheral proxy.} We wanted to see which one was more salient and allowed users to distinguish the target rendered on the peripheral display with greater confidence. But first, we aimed to empirically tune three parameters of the color enhancement algorithm so that the peripheral proxy was both salient in peripheral vision and visually consistent with the real-world target.

\begin{figure*}[!tbp]
  \centering
  \includegraphics[width=.6\textwidth]{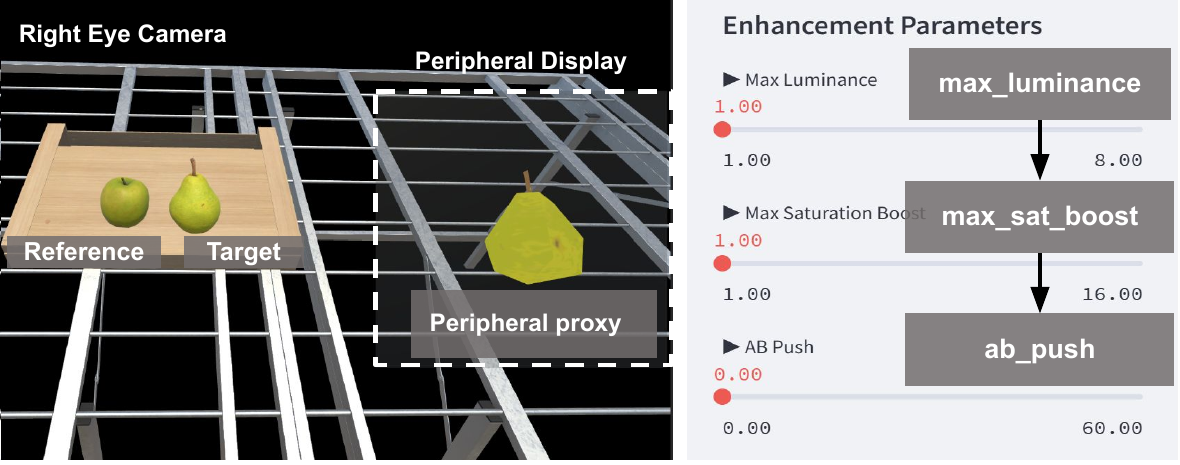}\vspace{-2ex}
  \caption{Calibration workflow. \textbf{Left:} Participants viewed a fruit shelf through left and right eye cameras; the peripheral proxy of the target fruit was rendered in the right-eye periphery. Participants assessed whether the proxy was visually consistent with the real-world target and distinguishable from the reference object. \textbf{Right:} We adjusted color enhancement parameters (\texttt{max\_luminance}, \texttt{max\_sat\_boost}, \texttt{ab\_push}) via a web interface not visible to participants. This participant--researcher loop iterated until the proxy appeared to be both salient in peripheral vision and consistent with the target, yielding empirically tuned parameter values.} \vspace{-2ex}
  \Description{The figure shows the calibration workflow for the color-enhancement system. On the left, a virtual fruit shelf is displayed with two objects labeled ``Reference'' (a green apple) and ``Target'' (a green pear). To the right of the target, a peripheral proxy of the pear is shown inside a dashed box labeled ``Peripheral Display.'' On the right side of the figure, a control panel interface is shown with three adjustable sliders labeled max_luminance, max_sat_boost, and ab_push, representing enhancement parameters. The layout illustrates how participants compared the proxy in peripheral vision with the in-world target while researchers adjusted parameters to calibrate the proxy's visibility and consistency.}
  \label{fig:calibration_workflow}
\end{figure*}

\p{Color Enhancement Parameter Calibration}
As illustrated in Fig.~\ref{fig:calibration_workflow}, each trial displayed a pair of fruits of similar color on a virtual shelf, with the left fruit serving as a reference and the right fruit as the target.
During a series of parameter adjustments, participants evaluated whether the peripheral proxy matched the target fruit while remaining noticeable with peripheral vision and distinguishable from the reference fruit only rendered real-world.

\begin{figure}[t]
  \centering
  \includegraphics[width=\linewidth]{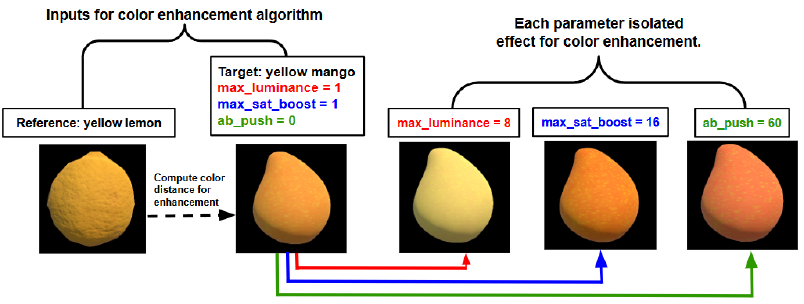}
  \vspace{-2ex}
  \caption{Each parameter's isolated effect for color enhancement. The target fruit (yellow mango) is enhanced relative to the reference fruit (yellow lemon). From left to right: baseline target, luminance boost, saturation boost, and chroma push. Each output shows the effect of a single parameter while the others remain fixed at their default values.}\vspace{-2ex}
  \Description{A comparison showing how different color enhancement parameters affect a yellow mango target relative to a yellow lemon reference. The left side shows the reference lemon and baseline mango with no enhancements. On the right, three versions of the mango are shown: a brighter version from increased luminance, a more saturated version from saturation boost, and a more color-shifted version from chroma push. Each output isolates the effect of one parameter while holding the others constant.}
  \label{fig:color_enhancement_variables}
\end{figure}

At the start of each trial, the peripheral proxy of the target fruit was rendered with default parameter settings. Similar to Study~1, participants first completed IPD fit and eye tracking calibration on Quest Pro to ensure accurate gaze tracking. We then fine-tuned the parameters sequentially using a dichotomous (binary search) procedure, i.e., for each parameter, the current color enhancement setting was compared against an alternative at the opposite bound of its range (\eg \texttt{max\_luminance}=1 \vs 8). Participants indicated which rendering viewed better for the peripheral proxy, defined as more salient and more differentiable from the reference fruit while still clearly matching the target fruit. The preferred value became the new baseline, and the parameter was then compared against the midpoint of the range. This halving procedure continued until a stable setting was reached, with at most three comparisons per parameter. The final value was then fixed before moving to the next parameter.
Three parameters were tuned (Fig.~\ref{fig:color_enhancement_variables}): 
\texttt{max\_luminance} (upper bound on luminance increase; default 1; range 1–8),
\texttt{max\_sat\_boost} (upper bound on saturation boost; default 1; range 1–16),
and \texttt{ab\_push} (Lab chromaticity shift magnitude; default 0; range 0–60).

In total, 8 participants (1 female, 7 male; ages 19--29, $M{=}24$, $SD{=}2.74$) with normal or corrected vision and no color‐vision deficiencies completed the calibration. All provided their informed consent and were compensated \$20 USD for their time.
%
%
We aggregated parameter settings across participants and reported the 75th percentile to derive conservative, perceptually effective bounds: \texttt{max\_luminance} = \textbf{2.125};
\texttt{max\_sat\_boost} = \textbf{9.75};
\texttt{ab\_push} = \textbf{30.0}. These group-level values reflect settings that ensure sufficient salience without over-enhancement.
We also report the grouped 75-percentile parameter values based on colors in Table~\ref{tab:calib_color_pct}.

\begin{table}[ht]
\vspace{-2ex}
\centering
\small
\begin{tabular}{lccc}
\toprule
\textbf{Color} & \textbf{Max Luminance} & \textbf{Max Saturation Boost} & \textbf{AB Push} \\
\midrule
Green  & 2.00   & 6.875   & 15.00  \\
Yellow & 3.00   & 10.50   & 38.00  \\
Red    & 3.00   & 10.50   & 28.75  \\
Blue   & 3.00   & 6.500   & 38.00  \\
\bottomrule
\end{tabular}
\caption{Group‐level 75th percentiles for each parameter by shelf color.}
\Description{Table with four rows and four columns showing group-level 75th percentile values of color enhancement parameters for different shelf colors. The first column lists the colors green, yellow, red, and blue. The next three columns provide the corresponding values for maximum luminance, maximum saturation boost, and AB push. Green has the lowest values overall, while yellow and red have higher saturation boost values. Blue has relatively high AB push but lower saturation boost.}
\label{tab:calib_color_pct}
\vspace{-8ex}
\end{table}

\section{Study 2: Peripheral Proxies}
\label{sec:study2}

Our final study applied the color enhancement calibration values to evaluate the \textit{screenshot} and \textit{Most Similar Color} (MSC) strategies for \textsc{PeriphAR} against a \textit{baseline} condition. The \textit{baseline} used a peripheral proxy with only color quantization---\ie without color enhancement---and applied the same decimation process. To maximize control and balance visual features, we simulated real-world targets using virtual objects with a video passthrough environment similar to Study~1; however, this time, the real-world objects were chosen as representations of real-world objects in a virtual fruit market scene.

\begin{figure*}[!tbp]
  \centering
  \includegraphics[width=.8\textwidth]{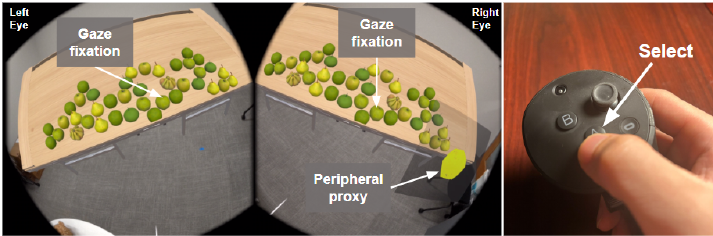}
  \vspace{-2ex}
  \caption{Study 2 setup. \textbf{Left:} Left/right eye views of the shelf; upon fixation on a target (\eg green apple), a peripheral proxy is rendered for confirmation. \textbf{Right:} The participant commits the selection with the “A” button; correct selections remove the gazed fruit.}
  \Description{Two side-by-side screenshots show the left and right eye views of a virtual shelf containing many green apples. A white arrow indicates gaze fixation on one apple, and in the right-eye view a yellow highlighted shape represents the peripheral proxy overlay. On the right side of the figure, a photo shows a hand holding a controller, with the ``A'' button labeled for making a selection.}
  \label{fig:system evaluation study scene}
  \vspace{-2ex}
\end{figure*}

\p{Task}
Each trial presented a virtual shelf with more than 40 similarly shaped and colored virtual fruits.
Participants selected two designated fruit types in a fixed alternating sequence (six of each; 12 correct selections total). They were instructed to make selections as accurately as possible without foveating toward the peripheral proxy, maintaining gaze on the shelf and using peripheral vision for confirmation.

\p{Procedure}
The study used a within-subjects design with three conditions (\textit{baseline}, \textit{screenshot}, \textit{MSC}) and three color-themed shelves (green, yellow, red fruits), yielding nine trials per participant. The set of fruit types associated with each shelf color (green, yellow, red), the positions of the designated fruits within each shelf, and the order of shelves were all fixed across participants. Shelf layouts were identical across conditions to maximize comparability. Only condition order was randomized, counterbalanced via a $3{\times}3$ Latin square (six groups; two participants per group). As in Study~1, participants first completed IPD fit and eye-tracking calibration to ensure accurate gaze tracking.

At the start of each trial, the two target fruit types were announced (\eg \emph{green apple} $\rightarrow$ \emph{guava} $\rightarrow$ \emph{green apple} $\rightarrow$ \emph{guava}, etc.). For each selection, the participant \textit{(1)} recalled the next fruit in sequence, \textit{(2)} fixated on the corresponding fruit on the shelf, \textit{(3)} received audio feedback as \textsc{PeriphAR} rendered the peripheral proxy for confirmation, and \textit{(4)} pressed a controller button to confirm the selection. Correct selections removed the target from the shelf with a confirmation sound; wrong selections left it visible with an error audio cue. After completing all three shelves for a condition, participants filled out a questionnaire and participated in a brief semi-structured interview before proceeding to the next condition. See Fig.~\ref{fig:system evaluation study scene} for the study setup.

\p{Participants}
In total, 12 participants (3 female, 9 male; ages 21--28, $M{=}24.7$, $SD{=}1.8$) with normal or corrected vision and no color-vision deficiencies took part. All provided informed consent and were compensated \$20 USD.

\p{Data Collection \& Analysis}
We logged gaze traces, gaze events, and selection events during the study session, together with post-questionnaire and interview feedback, providing objective and subjective measures.  
\changed{\textbf{Objective metrics} included: \textit{(1)} foveal \vs peripheral gaze distribution (percentage of gaze time spent looking directly at the peripheral proxy, lower is better), \textit{(2)} selection accuracy (proportion of correct selections, higher is better), and \textit{(3)} completion time (average time to complete one trial, lower is better).}  
\textbf{Subjective metrics} (7-point Likert scale, from post-condition questionnaire) included: confidence in selection, ease of noticing the peripheral proxy, perceived speed of confirmation, level of distraction (lower is better), perceived effectiveness, cognitive load (lower is better), reliance on color/shape cues of the peripheral proxy when selection, and gaze distribution (shelf vs.\ proxy, higher means more time on shelf). In addition, participants reported their \emph{overall preference} by ranking the three conditions.

We securely stored anonymized data and performed scripted preprocessing and statistical analysis. Similar to Study~1, ANOVA or Friedman significance tests, with Holm or Bonferroni corrections as appropriate, were applied for objective and subjective data analysis.

\p{Objective Performance}
\changed{Table~\ref{tab:objective_metrics} shows the objective metrics (gaze distribution, selection accuracy, completion time). We found no significant effect of condition per these metrics (all $p>0.05$).}

\begin{figure*}[ht]
  \centering
  \begin{minipage}[t]{0.30\textwidth}
    \centering
    \small
    \begin{tabular}{lccc}
    \toprule
    \textbf{Metric} & Base & SS & MSC \\
    \midrule
    Gaze (\%)  & 0.95 & 0.82 & 0.79 \\
    Error (\%) & 17.1 & 16.0 & 10.6 \\
    Time (s/trial)   & \changed{115.59}& \changed{91.90}& \changed{97.06} \\
    \bottomrule
    \end{tabular}
    \captionof{table}{Objective metrics.}
    \Description{Table showing objective metrics across three conditions (baseline, screenshot (SS), MSC). Metrics include gaze percentage, error percentage, and average time per trial in seconds. Baseline has the highest gaze percentage (0.95) but also higher error rate and longest time. MSC shows the lowest error rate (10.6\%) with shorter trial time compared to Baseline.}
    \label{tab:objective_metrics}
  \end{minipage}%
  \hfill
  \begin{minipage}[t]{0.28\textwidth}
    \centering
    \small
    \begin{tabular}{lrrr}
    \toprule
    Shelf & Wrong & Total & Err.\% \\
    \midrule
    Green   & 72  & 504 & 14.3 \\
    Yellow  & 123 & 555 & 22.2 \\
    Red     & 27  & 459 &  5.9 \\
    \bottomrule
    \end{tabular}
    \captionof{table}{Error rates by shelf.}
    \Description{Table showing error rates across three shelves (green, yellow, red). The yellow shelf has the highest error percentage (22.2\%), green is intermediate (14.3\%), and red shows the lowest error percentage (5.9\%).}
    \label{tab:shelf_overall}
  \end{minipage}%
  \hfill
  \begin{minipage}[t]{0.28\textwidth}
    \centering
    \small
    \begin{tabular}{lrrr}
    \toprule
    \bf Yellow & Wrong & Total & Err.\% \\
    \midrule
    Baseline   & 41 & 185 & 22.2 \\
    Screenshot & 51 & 195 & 26.2 \\
    MSC        & 31 & 175 & 17.7 \\
    \bottomrule
    \end{tabular}
    \captionof{table}{Yellow shelf error rates.}
    \Description{Table focusing on error rates for the Yellow shelf under three conditions (baseline, screenshot, MSC). The screenshot condition has the highest error rate (26.2\%), baseline is slightly lower (22.2\%), and MSC shows the lowest error rate (17.7\%).}
    \label{tab:yellow_shelf}
  \end{minipage}
  \vspace{-5ex}
\end{figure*}

To examine how the color context influenced errors, we aggregated error rates by shelf across all conditions (Table~\ref{tab:shelf_overall}). Because yellow shelves proved to be more error-prone than green and red shelves, we focused on the yellow shelf (Table ~\ref{tab:yellow_shelf}).
Interestingly, on the yellow shelf, the \textit{screenshot} condition produced a higher error rate (26.2\%) than \textit{baseline} (22.2\%), whereas \textit{MSC} reduced errors to 17.7\%.
We then analyzed the main distractors that the participants mistakenly selected (Table~\ref{tab:top_wrong_fruits}). In both \textit{baseline} and \textit{MSC}, errors were spread across yellow persimmon and Asian pear with similar frequency. In contrast, errors in the \textit{screenshot} condition were dominated by confusions with Asian pear (19 times), in the majority of cases when the intended target was \emph{yellow apple} (Table~\ref{tab:screenshot_confusions}).

\begin{figure*}[ht]
  \centering

  \begin{minipage}[t]{0.57\textwidth}
    \vspace{-1ex}
    \centering
    \scriptsize
    {\setlength{\tabcolsep}{4pt}\renewcommand{\arraystretch}{1.2}%
     \resizebox{\linewidth}{!}{%
      \begin{tabular}{l c c c}
        \toprule
        \textbf{Fruit Type} & \textbf{baseline} & \textbf{screenshot} & \textbf{MSC} \\
        \midrule
        Top 1 & yellow persimmon (13) & Asian pear (19)       & yellow persimmon (12) \\
        Top 2 & Asian pear (12)       & yellow mango (11)     & Asian pear (11)       \\
        Top 3 & apricot (6)           & yellow persimmon (10) & yellow mango (5)      \\
        \bottomrule
      \end{tabular}}}
    \captionof{table}{Top wrong fruit selections (yellow shelf).}
    \Description{A table showing the top three most frequent wrong fruit selections on the yellow shelf, broken down by condition (baseline, screenshot, MSC). In baseline, the top errors were yellow persimmon (13), Asian pear (12), and apricot (6). In screenshot, the top errors were Asian pear (19), yellow mango (11), and yellow persimmon (10). In MSC, the top errors were yellow persimmon (12), Asian pear (11), and yellow mango (5).}
    \label{tab:top_wrong_fruits}
  \end{minipage}%
  \hspace{0.015\textwidth} 
  \begin{minipage}[t]{0.41\textwidth}
    \vspace{0pt}
    \centering
    \scriptsize
    {\setlength{\tabcolsep}{5pt}\renewcommand{\arraystretch}{1.15}%
      \begin{tabular}{l r}
        \toprule
        \textbf{Intended fruit} & \textbf{Mis-selected as Asian pear} \\
        \midrule
        Yellow apple & 13 \\
        Yellow lemon & 6  \\
        \midrule
        \textbf{Total} & \textbf{19} \\
        \bottomrule
      \end{tabular}}
    \captionof{table}{Breakdown of Asian pear errors in the \textit{Screenshot} condition.}
    \Description{A breakdown of fruits that participants mis-selected as Asian pear in the screenshot condition. Yellow apple was mistaken 13 times, and yellow lemon was mistaken 6 times, totaling 19 errors.}
    \label{tab:screenshot_confusions}
  \end{minipage}
  \vspace{-5ex}
\end{figure*}

\changed{This pattern directly reflects how \textit{screenshot} constructs its reference image from a snapshot of the target on the yellow shelf with highly similar color distributions. The algorithm tries to enhance both the target and distractor but, in this case, against a very similar average, thus pushes in similar hue directions. This essentially reduces their perceptual distinctiveness in the peripheral display and results in insufficient enhancement.}
Participant feedback aligned with this interpretation. For example, P10 noted: \emph{``At the periphery, the Asian pear and yellow apple looked like uniform blocks of color, making them very difficult to differentiate [in \textit{screenshot}].''}

\begin{figure*}[ht]
  \centering

  \begin{minipage}[t]{0.48\textwidth}
    \centering
    \includegraphics[width=\linewidth]{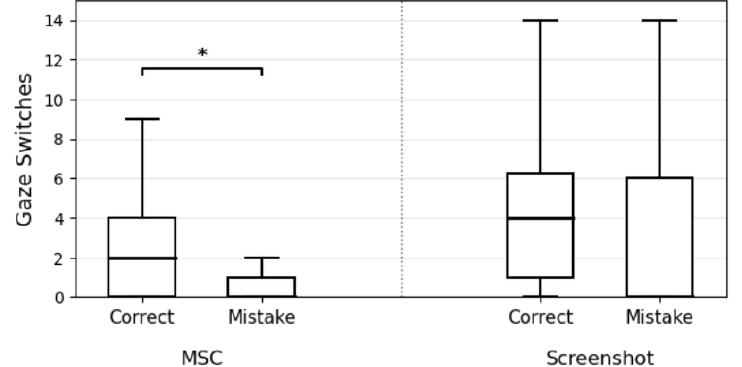}
    \captionof{figure}{\changed{Gaze switches between \emph{yellow apple} (target) and \emph{Asian pear} (distractor) on the yellow shelf for the \textit{MSC} and \textit{screenshot} conditions. Each box shows the distribution of gaze switches per trial grouped by condition and outcome (correct vs.\ mistake). The asterisk marks a significant difference between correct and mistaken selections in the \textit{MSC} condition (Mann--Whitney $U$ test, $p{=}0.016$); no significant difference was found for \textit{screenshot}.}}
    \Description{Box plot visualization of gaze switches between yellow apple (target) and Asian pear (distractor) on the yellow shelf, split by condition and outcome. Four boxes are shown in order: MSC--Correct, MSC--Mistake, Screenshot--Correct, and Screenshot--Mistake. The vertical axis is ``Gaze Switches.'' For MSC, correct trials have a noticeably higher distribution of gaze switches than mistakes, whose values cluster near zero. A horizontal bracket with an asterisk connects the MSC-Correct and MSC-Mistake boxes, indicating a significant difference. For Screenshot, the correct and mistake boxes overlap more, with no significance marker. The figure emphasizes that in difficult trials, making additional comparisons (more gaze switches) helped when using MSC, but not when using Screenshot.}
    \label{fig:yellow_apple_asian_pear_gaze_switches}
  \end{minipage}
  \hfill
  \begin{minipage}[t]{0.48\textwidth}
    \centering
    \includegraphics[width=\linewidth]{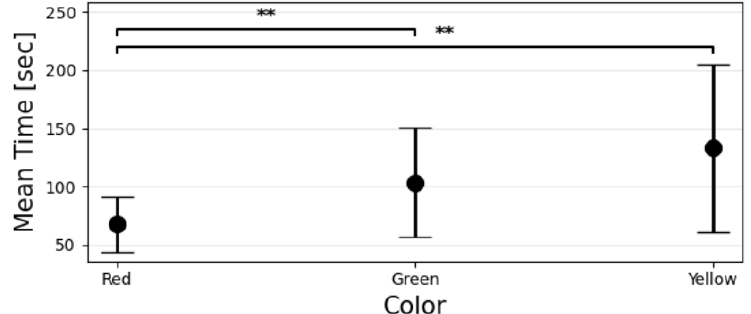}
    \captionof{figure}{\changed{Mean completion times across target colors in Study~2. Error bars show $\pm$\,SD. Asterisks indicate significant pairwise differences based on a Friedman test with Wilcoxon signed-rank post-hoc comparisons (Holm-corrected).}}
    \paragraph{Dot plot showing mean completion times for Study 2 trials across three target color shelves: red, green, and yellow. Each color category is represented by a bar or dot with vertical error bars showing standard deviation. The red shelf has the shortest mean time, green is slower, and yellow is slowest overall. Asterisks above the red--green and red--yellow pairs indicate significant differences, while no asterisk appears between green and yellow. The figure conveys that red targets were completed significantly faster than green and yellow, consistent with preattentive advantages for red.}
    \label{fig:study_2_completion_time_across_color}
  \end{minipage}

\end{figure*}

In contrast, \textit{MSC} explicitly selects the neighbor with the most similar color as reference. For the yellow shelf, \emph{yellow apple} and \emph{Asian pear} were the closest pair in \textsf{CIELAB} space. When \emph{Asian pear} was the target, \emph{yellow apple} became the reference, and the algorithm enhanced the pear proxy to increase contrast. Conversely, when \emph{yellow apple} was the target and \emph{Asian pear} the reference, the apple was darker and therefore left unenhanced by the algorithm.
This asymmetry in the algorithm produced a visible difference in the periphery.
In post-condition interviews, five participants reported deliberately alternating their gaze between the two fruits to exploit this difference, noting that \textit{MSC} made such comparisons more reliable. \changed{Across the study, most correct selections did \emph{not} involve such deliberate comparisons: depending on shelf and condition, roughly half to two thirds of correct trials were completed with zero gaze switches between target and distractors (e.g., $57\%$ for \textit{baseline}, $61\%$ for \textit{screenshot}, and $64\%$ for \textit{MSC} across all shelves). This matches our design goal that PeriphAR should normally support ``one-look'' confirmation. Gaze switching therefore mainly occurred in a subset of more difficult trials where participants chose to double-check ambiguous proxies.} \changed{To understand these harder cases, we examined gaze switches between \emph{yellow apple} (target) and \emph{Asian pear} (primary distractor) on the yellow shelf, where errors were most concentrated.
In the \textit{screenshot} condition, \changed{participants made 13 mis-selections}. When they incorrectly selected the \emph{Asian pear}, they performed on average $3.64$ gaze switches ($SD{=}4.96$), compared to $7.11$ ($SD{=}11.45$) for correct selections. This difference was not significant (Mann--Whitney $U$, $p{=}0.097$), suggesting that additional comparisons between the two peripheral proxies did not reliably resolve their ambiguity.
In the \textit{MSC} condition, participants made only 7 mis-selections, and they were typically made with almost no comparison ($M{=}0.57$, $SD{=}0.90$), whereas correct selections involved some back-and-forth gaze switches ($M{=}4.64$, $SD{=}9.24$). These differences were significant (Mann--Whitney $U$, $p{=}0.016$).
Together with the lower error rate on the yellow shelf for \textit{MSC}, this pattern indicates that when participants chose to compare \emph{yellow apple} and \emph{Asian pear}, the asymmetric enhancement strategy in \textit{MSC} produced more diagnostic peripheral differences, whereas the scene-wide reference in \textit{screenshot} often left the two compared fruits (target and primary distractor) perceptually similar even after additional gaze switches to compare the intended target and the shown peripheral proxy (Fig.~\ref{fig:yellow_apple_asian_pear_gaze_switches}). Importantly, many trials in all conditions were completed without any gaze switching, so \textsc{PeriphAR} is not designed to rely on deliberate back-and-forth inspection; rather, \textit{MSC} appears to offer more useful information in those difficult comparisons.}

\changed{We also examined whether target color affected completion time. The red fruit shelf were completed fastest ($M{=}67.78$\,s, $SD{=}36.64$), followed by green ($M{=}103.90$\,s, $SD{=}61.29$) and yellow ($M{=}132.98$\,s, $SD{=}100.20$). A Friedman test revealed a significant main effect of color, $\chi^2(2){=}12.67$, $p{=}0.0018$. Post-hoc Wilcoxon signed-rank tests with Holm correction showed that red trials were significantly faster than both yellow ($p_{\mathrm{holm}}{=}0.0044$) and green ($p_{\mathrm{holm}}{=}0.0098$), whereas time differences for green and yellow were not  significant~(Fig.~\ref{fig:study_2_completion_time_across_color}).
The significant differences for red align with prior work on preattentive color processing, which reports that red tends to guide attention more efficiently than other hues~\cite{andersen2019attentional,nothdurft1993preattentive}.}

\begin{figure*}[t]
  \centering

  \begin{minipage}[t]{0.44\textwidth}
    \vspace{0pt}
    \centering
    \scriptsize
    \begin{tabular}{lccc}
      \toprule
      \textbf{Metric} & \textbf{baseline} & \textbf{screenshot} & \textbf{MSC} \\
      \midrule
      Color cue reliance        & 5.667 & 5.250 & 5.500 \\
      Shape cue reliance        & 5.083 & 4.583 & 5.417 \\
      \addlinespace
      \textbf{Confidence}\textsuperscript{*}          & 4.833 & 5.167 & 6.167 \\
      \textbf{Ease of noticing}\textsuperscript{*}    & 5.000 & 5.667 & 6.250 \\
      \addlinespace
      Speed of confirmation      & 5.000 & 5.250 & 5.417 \\
      Distraction (lower, better)        & 1.917 & 1.750 & 1.583 \\
      Effectiveness          & 4.833 & 5.083 & 5.500 \\
      Gaze distribution             & 4.833 & 5.250 & 5.833 \\
      Cognitive load (lower, better)     & 3.000 & 2.500 & 2.167 \\
      \bottomrule
    \end{tabular}

    \vspace{0.25ex}
    \noindent\scriptsize\textsuperscript{*}\,Significant difference ($p{<}0.05$).

    \captionof{table}{Mean subjective ratings (1--7). Higher is better unless noted otherwise.}
    \Description{Table comparing subjective ratings (scale 1--7) across three conditions: baseline, screenshot, and MSC. Metrics include color cue reliance, shape cue reliance, confidence, ease of noticing, speed of confirmation, distraction, effectiveness, gaze distribution, and cognitive load. Higher values generally indicate better performance, except for distraction and cognitive load where lower is better. Confidence and ease of noticing show significant differences, with MSC achieving the highest ratings on both. MSC also has the lowest distraction and cognitive load scores.}
    \label{tab:subjective_all_metrics}
    \vspace{-6ex}
  \end{minipage}
  \hfill
  \begin{minipage}[t]{0.55\textwidth}
    \vspace{0pt}
    \centering
    \includegraphics[width=\linewidth]{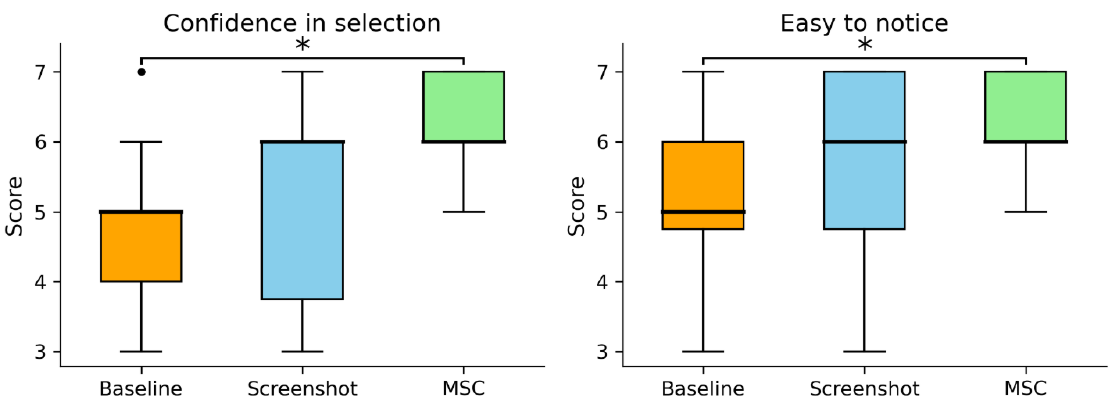}
    \vspace{-2ex}
    \captionof{figure}{Box plots for \emph{Confidence in selection} and \emph{Ease of noticing}. Asterisks mark significant pairwise differences (Holm-corrected).}
    \Description{Two box plots compare baseline, screenshot, and MSC conditions on subjective ratings. The left plot shows confidence in selection; the right plot shows ease of noticing the proxy. For both plots, each condition is represented by a separate box with whiskers indicating score ranges. MSC has the highest median confidence and ease-of-noticing scores, baseline has the lowest, and Screenshot is in between. Asterisks above certain pairs mark statistically significant differences (\eg MSC \vs~baseline). The figure highlights that participants felt more confident and found the proxy easier to notice with MSC.}
    \label{fig:subjective_significant_box_plot}
  \end{minipage}
  \vspace{-1ex}
\end{figure*}

\p{Subjective Performance}
Table~\ref{tab:subjective_all_metrics} shows subjective ratings from the post-condition questionnaires. Overall, the scores tended to favor conditions with color enhancement, particularly using the \textit{MSC} strategy. We found significant differences for two key measures (Fig.~\ref{fig:subjective_significant_box_plot}): \textit{confidence in selection} (Friedman $\chi^2(2){=}9.15$, $p{=}0.010$; \textit{MSC} $>$ \textit{baseline} $p_{\mathrm{holm}}{=}0.022$) and \textit{ease of noticing the proxy} (Friedman $\chi^2(2){=}11.89$, $p{=}0.003$; \textit{MSC} $>$ \textit{baseline}, $p_{\mathrm{holm}}{=}0.020$). \changed{Participants also reported relying slightly more on shape information in the peripheral proxy under \textit{MSC} (M{=}5.42) than under \textit{baseline} (M{=}5.08) or \textit{screenshot} (M{=}4.58; Table~\ref{tab:subjective_all_metrics}), but these differences were not statistically significant.
This trend can be explained by considering that enhancing color relative to the most similar colored neighbor can increase local chromatic contrast along the target's edges. As a result, \textit{MSC}'s enhancement may sometimes have made object boundaries in the peripheral proxy easier to interpret. This is consistent with work showing that preattentive processing of color can facilitate segmentation and grouping~\cite{nothdurft1993preattentive,andersen2019attentional}. Qualitative interview feedback echoed this interpretation: P10 noted, \emph{``The shape was more noticeable, and I could discern the fruit's stamp direction in the peripheral cue, which helped me differentiate between similar fruits.''}, and P2 felt that \emph{``the shape was clearer in this [\textit{MSC}] condition, which helped [them] make decisions beyond just color.''}}

The \textit{MSC} condition was chosen by 8 of 12 participants, \textit{screenshot} by 3, and \textit{baseline} by 1. A chi-square test against uniform preference was significant, $\chi^2(2){=}6.18$, $p{=}0.045$. Pairwise two-proportion $z$-tests with Holm correction revealed a significant preference for \textit{MSC} over \textit{baseline} ($z{=}2.951$, $p_{\mathrm{holm}}{=}0.009$).

\p{Reflection \& Takeaways}

In summary, the \textit{MSC} condition improved the subjective experience, leading to higher preference for the condition, selection confidence, and peripheral noticeability over the baseline.
\changed{While there was no significant effect of condition per the objective metrics, deeper analysis showed trends in favor of \textit{MSC}.} Our error selection analysis revealed that color context mattered: on yellow shelves, \textit{screenshot} amplified confusion with similarly colored fruits (Asian pear), whereas \textit{MSC} was most robust across shelves. The \textit{screenshot}'s scene‐wide reference tended to homogenize color enhancement across similar objects, inadvertently amplifying confusion in cluttered contexts. Conversely, the \textit{MSC} strategy leveraged the most similar neighbor to accentuate subtle but critical differences, yielding more reliable peripheral confirmation in visually challenging scenarios, notably the yellow shelf trials.
Participant feedback suggests that color enhancement relative to the most similar object in color goes beyond improving color contrast alone, potentially strengthening peripheral shape perception as well.
\changed{Finally, our analysis of completion times by target color suggests that PeriphAR inherits classic preattentive color patterns: red targets were noticed fastest, with green and yellow slower, mirroring established findings on color-guided attention~\cite{andersen2019attentional,nothdurft1993preattentive}.}

\section{End-to-end System}
\label{sec:end-to-end}

\changed{As a final step, we wanted to see how \textsc{PeriphAR} might perform in less controlled environments compared to our lab studies. However, we don't think of this step as a generalization attempt; rather, we simply wanted to learn if and how it could be made more effective when part of an end-to-end system and tested in potentially challenging settings with many similarly looking, cluttered physical objects (vending machines, bookshelves, etc.).
To enable this exploration, we extended the system with real-time support for \textit{(1) processing live camera video passthrough} and \textit{(2) object detection and segmentation}, then profiled each step in the pipeline to identify where most computing resources might be needed.}

\p{Switching to Real-World Objects}

First, accessing the externally facing cameras on Quest Pro is an experimental feature. The recently released Passthrough API is only available for Quest 3. We were able to exploit an Android media projection workaround to screen-grab the live video feed. Next, we needed to know where the user is looking within the passthrough to isolate the target object. However, converting the eye gaze cursor from 3D coordinates to the corresponding point in the 2D passthrough image was challenging because of barrel distortion and the eye gaze originating from the left eye alone. As a workaround, we added a red dot to indicate where the user is looking, so that the resulting 2D image also contains this red dot in the accurate location. The system can then parse the image for the center of the reddest pixels to find the 2D location of the user's gaze.

Second, when transitioning from virtual fruits to real-world objects, we needed object segmentation to isolate individual objects and grab their corresponding masks. The YOLO11n segmentation model\changed{~\cite{khanam2024yolov11}} was used to maximize accuracy while simultaneously limiting latency. Once the objects are segmented, the system determines if the decoded gaze pixel resides within any of the detected object masks and if so, that object is treated as the target. The bounding box for that target is expanded in all directions to create a 2D box collider. If any other object bounding boxes intersect this expanded box, then it is treated as a neighboring object. Once the target mask and neighbor masks are gathered, the system runs the color similarity algorithm to determine which neighbor mask is most similar in color to the target mask. Once the target and reference masks are determined, a color-enhanced and quantized version of the target object is created. The enhanced image is then finally displayed in the simulated peripheral display.

\begin{figure}
    \centering
    \includegraphics[width=\linewidth]{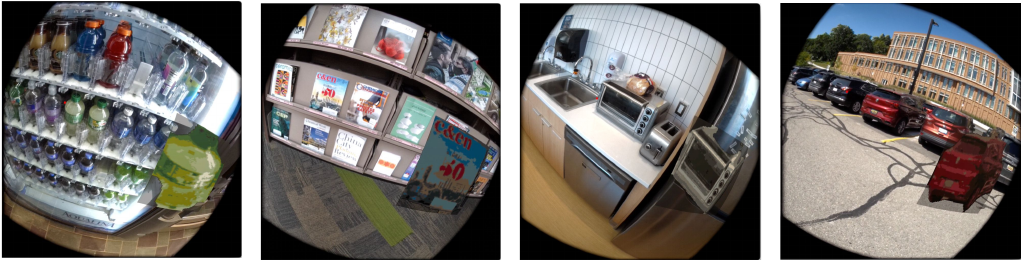}
    \caption{Examples from our end-to-end system tests showing only the right eye view with peripheral proxy on simulated peripheral display.}
    \Description{A set of camera images showing examples from the end-to-end system tests, each presenting the right-eye view with a peripheral proxy overlay on the simulated peripheral display. Scenes include: A vending machine with drink bottles, where a bright proxy marks the selected beverage. A library with book shelves, with a proxy highlighting one magazine or book. A kitchen counter with appliances such as an oven or microwave, with a proxy over a control panel or handle. An outdoor parking lot with a car, where the proxy highlights the car to be found. In each image, the proxy appears near the edge of the field of view, floating over the corresponding real-world object, illustrating how PeriphAR supports peripheral confirmation in everyday tasks.}
    \label{fig:realworld}
\end{figure}

\p{Testing in Realistic Settings}
We then used the resulting end-to-end system for a variety of real-world tasks (Figure~\ref{fig:realworld}): buying a drink at a vending machine, selecting magazines and books in a library, operating a kitchen appliance, and finding a car.
\changed{We treat these deployments as proof-of-concept demonstrations, intended to show that the full PeriphAR pipeline can run on current XR hardware and to obtain failure cases, rather than as a comprehensive evaluation of deployment-ready performance.}
\changed{We chose to not incorporate a fixation threshold on the end-to-end system, unlike the previous two studies, as applying object segmentation to the real world already takes a significant amount of time and acts as a threshold (min.~time of 113ms) ensuring the user is indeed viewing a relevant object. Accompanied with the fact that per-frame fixation is observed on the server-side, as opposed to the headset, we decided that a fixation threshold would unnecessarily increase the latency between viewing a physical object and receiving the enhanced mask. Similarly, request rates between the headset and server were limited as to not overload the network.}

In these examples, going through the entire process from acquiring the passthrough image with the real-world target to rendering the final color enhanced target took on average between 0.72s (parking lot, where enhancement was often skipped due to no other objects being detected) and 1.55s (kitchen appliance, with an average of 2.6 neighbors including refrigerator, sink, chair, bowl, table, oven).
During profiling the various steps of the pipeline, we found that the object segmentation takes on average about 25\% of the time, while the mean time to find the most similar neighbor object ranged from 17.8\% (vending machine, avg. neighbors 15.6 including refrigerator, bottle) to 28.5\% of the time (\changed{parking lot, avg. neighbors 2.3--3.9 including bus, car, truck, person}) depending on the complexity of the scene and size of neighbors. Lastly, the quantization and color enhancement of the image typically took a majority of the time ranging from 50.1\% (vending machine) to 61.9\% (kitchen). 

\changed{Beyond latency, two practical aspects are important for interpreting these deployments: occlusion safety and lighting conditions. First, in the headset, the monocular peripheral display occupies only a small $\sim$20$^{\circ}$ horizontal FOV and is rendered on a semi-transparent plane, leaving the central view unobstructed and limiting occlusion in the periphery.\footnote{See Section~\ref{sec:simulation} for details on display emulation.} Although the proxy appears larger in the captured video frames in Fig.~\ref{fig:realworld}, its apparent size in the headset is quite modest. In our end-to-end system use cases, participants used \textsc{PeriphAR} while standing still or moving slowly, and the proxy was shown only in brief bursts (2\,seconds) to confirm a selection. Deploying such overlays in safety-critical situations (e.g., approaching traffic or navigating uneven terrain) would require additional safeguards, such as more conservative opacity and size, or temporarily suppressing the peripheral proxy during rapid locomotion.}

\changed{Second, because the peripheral proxy is rendered directly into the passthrough view, its visibility depends on both the headset display and the captured real-world lighting condition. In our tests, the end-to-end system was used mainly in indoor environments and in typical daylight outdoors (sunny but not high-glare conditions), where the Quest Pro display provided sufficient dynamic range for the proxy to remain clearly visible. We did not systematically vary ambient lighting, and in very bright sunlight or strong glare the same color enhancement would have less headroom on the physical display, reducing perceived contrast between proxy and background. A more lighting-aware version of \textsc{PeriphAR} that adapts enhancement strength to ambient conditions is therefore an important direction for future work.}

\begin{figure*}[!tbp]
  \centering
  \includegraphics[width=\textwidth,height=0.40\textheight,keepaspectratio]{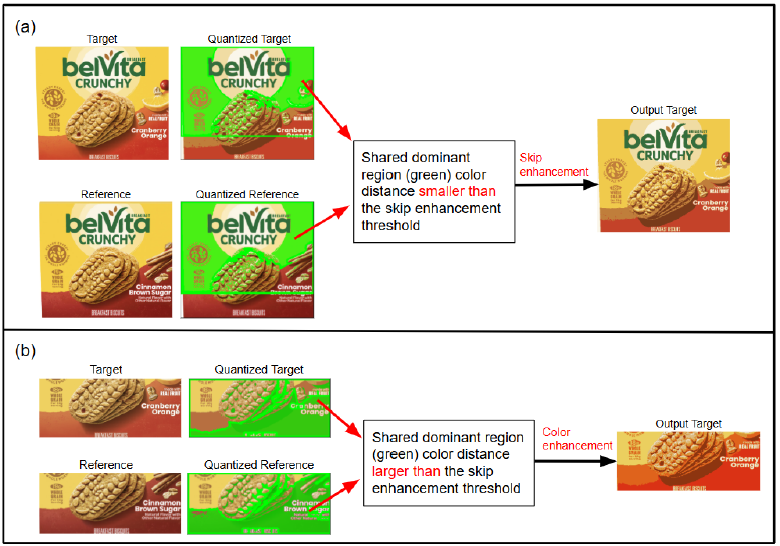}
  \vspace{-2ex}
  \caption{Effect of the color‐enhancement algorithm on real‐world objects with multiple color regions, illustrated using two \textit{BelVita} cookie packages. 
\textbf{(a)} With full packages, the algorithm selects the shared yellow header as the dominant region. Since the color distance of the shared dominant region between target and reference is below the enhancement threshold, enhancement is skipped, and the target proxies remain visually similar to the reference. 
\textbf{(b)} With only the lower halves, the dominant regions (orange \vs~red) differ more, exceeding the threshold and triggering enhancement, making the target proxy more distinguishable to the reference.}
  \Description{The figure has two panels labeled (a) and (b), each showing a sequence of images demonstrating how the algorithm processes BelVita cookie packages. Panel (a): The top row shows the target package (Cranberry Orange) next to its quantized version, which highlights the yellow header region in green. The bottom row shows the reference package (Cinnamon Brown Sugar) and its quantized version, also dominated by green in the header region. Red arrows point from the quantized images to a central box labeled ``Shared dominant region (green) color distance smaller than the skip enhancement threshold.'' A black arrow labeled ``Skip enhancement'' then points to the final output target on the right, which looks nearly identical to the original target package. Panel (b): The same layout is used but with cropped lower halves of the packages. The target’s lower region appears orange, while the reference’s lower region is red. Their quantized versions highlight these areas in green masks. Red arrows point to a central box labeled ``Shared dominant region (green) color distance larger than the skip enhancement threshold.'' A red arrow labeled ``Color enhancement'' leads to the final output target on the right, where the orange region has been visibly boosted, making the package stand out more clearly from the red reference.
}
  \label{fig:belvita}
  \vspace{-3ex}
\end{figure*}

\p{Limitations in Real-World Scenarios}
While our algorithm performed reliably in controlled conditions, its behavior with real-world objects that contain multiple prominent color regions reveals a notable limitation. Consider the example of two \textit{BelVita} cookie packages, Cranberry Orange (target) and Cinnamon Brown Sugar (reference), shown in Fig.~\ref{fig:belvita}. Both share a large yellow header region at the top of the package, while their lower halves differ in dominant hue (orange \vs~red). When the color enhancement algorithm processes the full packages, it identifies the shared yellow header as the dominant overlapping region (see Fig.~\ref{fig:belvita}a). Because this region is nearly identical across both packages, the similarity measure falls below the skip enhancement threshold. This threshold serves as a safeguard: when two textures are already visually close, additional enhancement would either have little perceptual benefit or risk distorting the target's appearance. By skipping enhancement in such cases, the algorithm avoids unnecessary adjustments and preserves alignment with the real-world object. As a result, the peripheral proxies remain visually similar.


This mismatch highlights an important limitation: the algorithm assumes that the most shared dominant region is the critical region for differentiation. One way to address this is to redirect the color enhancement algorithm toward the subdominant but visually distinguishing regions when dominant regions are nearly identical. As an experiment, extracting and processing only the lower halves of the packages led the algorithm to enhance successfully (see Fig.~\ref{fig:belvita}b), underscoring the gap between algorithmic focus and human perceptual strategies. 
\changed{Our exploratory analysis of Study~2 completion times (Fig.~\ref{fig:study_2_completion_time_across_color}) also suggests that not all colors contribute equally to fast peripheral detection: red targets were noticed significantly faster than green and yellow, echoing preattentive color findings~\cite{andersen2019attentional,clifford2010color}. Taken together with the BelVita example, this points to a perceptually informed extension of our MSC strategy from object-level to patch-level enhancement. Rather than treating each object as a single texture, the system could segment it into chromatic subregions and, for each subregion, estimate \textit{(i)} how strongly it differentiates the target from its most similar color neighbor (i.e., the color difference between the corresponding areas in target and neighbor, discounting large regions that are nearly identical such as the shared yellow header), and \textit{(ii)} its predicted preattentive salience based on hue category (e.g., reddish/orangish accents tend to guide attention more strongly). Color enhancement could then be concentrated on subregions that both show strong target--neighbor contrast and are high-salience, instead of boosting the entire object uniformly. Such a patch-wise extension would preserve the current MSC mechanism of choosing the most similar color neighbor as reference, while making PeriphAR more robust for multi-colored, cluttered scenes (e.g., packaged goods), where users often rely on small distinctive regions in the periphery to disambiguate similar items. We view our BelVita case as highlighting where the present proof-of-concept starts to break down and as motivating these perceptually informed refinements as concrete future work toward improved generalizability in real-world deployments.}

\changed{More broadly, our prototype should be interpreted as a proof-of-concept for peripheral confirmation on low-FOV AR glasses rather than a deployment-ready system. Our end-to-end system focused on relatively low-risk scenarios (e.g., standing or slow movement), not fast locomotion or safety-critical tasks, so any real-world deployment would need to incorporate explicit safety constraints on when and where peripheral proxies may appear.}

\section{Discussion}
\label{sec:discussion}


\p{Opportunities \& Tradeoffs with Display Simulation}
The system we created to design our \textsc{PeriphAR} technique provided a \textit{quasi-realistic} simulation of future AR glasses on Quest Pro, enabling peripheral vision experiments in the absence of wide-FOV AR displays. Simulation is an important tool in HCI research to experience advanced technology \cite{Murray-Smith-Interactions22}. We posit that future AR glasses will have to be limited in critical ways and, even if more capable with larger FOVs, users may not always want full-on AR. So, our idea of adopting \textit{graceful degradation} as a design principle could provide a way forward.

While we believe in the practical value and applicability of our work, findings are based on simulation. We modeled display characteristics like position and FOV digitally. Without the hardware we cannot verify accuracy. Another critical aspect is that our controlled setting was free from peripheral distractions in the real world. With optical see-through displays, digital eye strain \cite{Hirzle-ETRA20} will likely be more pronounced. In our studies, eye movements when switching from real-world targets to the simulated display required relatively little accommodation. Using a unifocal video passthrough display meant that the eyes' focal length did not need to adjust as much.

\p{Context-Aware Parameter Tuning}
While we selected a single global setting for our color-enhancement algorithm parameters based on group-level 75th-percentile values, the calibration results indicated distinct optimal values for different colors. For instance, participants generally required less aggressive enhancement for green fruits compared to yellow or red fruits (see Table~\ref{tab:calib_color_pct}). Consequently, participants with higher peripheral sensitivity occasionally found certain peripheral proxies to be overly enhanced. For example, the green apple proxy sometimes made alignment with the real-world object challenging. Future implementations should adopt context-aware parameter tuning, dynamically adjusting enhancement parameter values based on color context to ensure consistent peripheral proxy effectiveness across diverse visual scenarios.

\p{Neighbor-Aware Enhancement Bounds} 
In the current \textit{MSC} implementation, enhancement is based solely on the single adjacent object with the closest color. For example, in a green shelf trial, when participants fixated on a lighter green apple, the algorithm selected a darker green guava as the reference. This shifted the  peripheral proxy of the apple brighter toward yellow-green, creating a stronger contrast with the green guava. However, this also risked enhancing the proxy to the point that it no longer aligned well with the green apple, instead shifting toward a lighter, more yellowish green that resembled the nearby green pear. This over-enhancement made it harder for participants to match the proxy to the intended target in peripheral vision. To address this limitation, future work should extend the \textit{MSC} strategy with a neighbor-aware constraint mechanism. After selecting the most similar neighbor, \textsc{PeriphAR} could check all other adjacent objects and clamp enhancement so that the target never exceeds any neighbor. In this way, enhancement would still increase target–neighbor contrast but remain bounded by the broader visual context. Such safeguards would reduce the risk of overly bright or misleading proxies and better preserve perceptual alignment in complex, multi-object environments. These refinements highlight promising directions for improving the robustness and generalizability of peripheral color enhancement in AR-based selection tasks.


\p{Loss of Fine-Grained Details}
Although color enhancement increases overall salience, it can inadvertently obscure subtle features that aid recognition. For example, P12—the only individual who preferred the \textit{baseline} condition—reported difficulty aligning the peripheral proxy with the real-world target on the fruit shelf. In the \textit{baseline} condition, the proxy is only quantized, which preserves the dark striping on the tomato's surface. After enhancement in other two conditions, however, those stripes blended into a uniform red, making the peripheral proxy appear “too flat” and harder to match to the real fruit. Future refinements should balance global color boosts with the preservation of local contrast and edge details. This would help maintain both visibility and recognizable texture cues.

\p{Simpler Shape Representations}
As our selection error analysis in Study~1 showed, targets close in hue and shapes with only minor differences led to the most errors.
The decimation step prior to color enhancement in Study~2 mitigated these issues; however, color enhancement was based on the full textures of the virtual fruits rather than only on the viewable perspective.
Our end-to-end system used object masks from the segmentation based on low-res passthrough images, skipping 3D representation altogether making visual similarity dependent on viewing angle and size of region.
While modern segmentation and 3D reconstruction techniques such as \textit{Segment Anything Model} (SAM) \cite{Kirillov2023} or \textit{Neural Radiance Fields} (NeRF) \cite{Mildenhall-ECCV2020} could perhaps be added one day, such fine-grained segmentation and photo-realistic reconstruction would be overkill.
Our work calls for new techniques that abstract real-world objects by compressing their texture and shape into simple yet distinguishable visual patterns.
For example, finding rough visual contours approximating an object to its primitive shape yet exaggerating certain features in the context of nearby objects, or semantically shifting colors based on keys that humans associate with particular objects, such as green for apples or yellow for bananas despite not being plain in color, could further improve glanceability.

\section{Conclusion}

This paper systematically developed \textsc{PeriphAR}, a gaze-based feedback technique that recreates real-world targets as peripheral proxies optimized for glanceability. To do this, we first developed the key technical components to simulate future AR glasses with peripheral display modes operated in monocular fashion on Quest Pro using graceful degradation. We demonstrated that fundamental tasks like object selection can be performed efficiently and accurately with our AR display designs that used color enhancement to become peripherally glanceable.
We discussed the limitations of our display simulation approach and identified opportunities for improvement of our technique, pointing toward a roadmap for advancing \textsc{PeriphAR} toward real deployment in future always-on AR glasses.
By shifting confirmation feedback to the user's peripheral field, our work opens a new design space for lightweight, low-powered AR glasses that align with both hardware constraints and human perceptual strengths.

\bibliographystyle{ACM-Reference-Format}
\bibliography{bib/xrstudio.bib,bib/xrcam.bib,bib/grandvalley,bib/periphar}

@inproceedings{Mousavian-CVPR17,
  title={3D Bounding Box Estimation using Deep Learning and Geometry},
  author={Mousavian, Arsalan and Anguelov, Dragomir and Flynn, John and Kosecka, Jana},
  booktitle={Proc.~CVPR},
  pages={7074--7082},
  year={2017}
}

@inproceedings{Mildenhall-ECCV2020,
  title={{NeRF: Representing Scenes as Neural Radiance Fields for View Synthesis}},
  author={Mildenhall, B and Srinivasan, PP and Tancik, M and Barron, JT and Ramamoorthi, R and Ng, R},
  booktitle={Proc.~ECCV},
  year={2020}
}

@inproceedings{Bergstrom-CHI21,
  author       = {Joanna Bergstr{\"{o}}m and
                  Tor{-}Salve Dalsgaard and
                  Jason Alexander and
                  Kasper Hornb{\ae}k},
  editor       = {Yoshifumi Kitamura and
                  Aaron Quigley and
                  Katherine Isbister and
                  Takeo Igarashi and
                  Pernille Bj{\o}rn and
                  Steven Mark Drucker},
  title        = {How to Evaluate Object Selection and Manipulation in VR? Guidelines
                  from 20 Years of Studies},
  booktitle    = {Proc.~CHI},
  pages        = {533:1--533:20},
  publisher    = {{ACM}},
  year         = {2021},
  url          = {https://doi.org/10.1145/3411764.3445193},
  doi          = {10.1145/3411764.3445193},
  bibsource    = {dblp computer science bibliography, https://dblp.org}
}

@inproceedings{Luyten-CHI16,
  author       = {Kris Luyten and
                  Donald Degraen and
                  Gustavo Alberto Rovelo Ruiz and
                  Sven Coppers and
                  Davy Vanacken},
  editor       = {Jofish Kaye and
                  Allison Druin and
                  Cliff Lampe and
                  Dan Morris and
                  Juan Pablo Hourcade},
  title        = {Hidden in Plain Sight: an Exploration of a Visual Language for Near-Eye
                  Out-of-Focus Displays in the Peripheral View},
  booktitle    = {Proc.~CHI},
  pages        = {487--497},
  publisher    = {{ACM}},
  year         = {2016},
  url          = {https://doi.org/10.1145/2858036.2858339},
  doi          = {10.1145/2858036.2858339},
  bibsource    = {dblp computer science bibliography, https://dblp.org}
}

@inproceedings{Sendhilnathan-UIST22,
  author       = {Naveen Sendhilnathan and
                  Ting Zhang and
                  Ben Lafreniere and
                  Tovi Grossman and
                  Tanya R. Jonker},
  editor       = {Maneesh Agrawala and
                  Jacob O. Wobbrock and
                  Eytan Adar and
                  Vidya Setlur},
  title        = {Detecting Input Recognition Errors and User Errors using Gaze Dynamics
                  in Virtual Reality},
  booktitle    = {Proc.~UIST},
  pages        = {38:1--38:19},
  publisher    = {{ACM}},
  year         = {2022},
  url          = {https://doi.org/10.1145/3526113.3545628},
  doi          = {10.1145/3526113.3545628},
  bibsource    = {dblp computer science bibliography, https://dblp.org}
}

@article{Pfeuffer-CG21,
  title={ARtention: A design space for gaze-adaptive user interfaces in augmented reality},
  author={Pfeuffer, Ken and Abdrabou, Yasmeen and Esteves, Augusto and Rivu, Radiah and Abdelrahman, Yomna and Meitner, Stefanie and Saadi, Amr and Alt, Florian},
  journal={Computers \& Graphics},
  volume={95},
  pages={1--12},
  year={2021},
  publisher={Elsevier}
}

@article{Liang-CG94,
title = {JDCAD: A highly interactive 3D modeling system},
author = {Jiandong Liang and Mark Green},
journal = {Computers \& Graphics},
volume = {18},
number = {4},
pages = {499-506},
year = {1994},
issn = {0097-8493},
doi = {https://doi.org/10.1016/0097-8493(94)90062-0},
url = {https://www.sciencedirect.com/science/article/pii/0097849394900620}
}

@inproceedings{Costanza-MobileHCI06,
  author       = {Enrico Costanza and
                  Samuel A. Inverso and
                  Elan Pavlov and
                  Rebecca Allen and
                  Pattie Maes},
  editor       = {Marko Nieminen and
                  Mika R{\"{o}}ykkee},
  title        = {eye-q: eyeglass peripheral display for subtle intimate notifications},
  booktitle    = {Proc.~Mobile HCI},
  pages        = {211--218},
  publisher    = {{ACM}},
  year         = {2006},
  url          = {https://doi.org/10.1145/1152215.1152261},
  doi          = {10.1145/1152215.1152261},
  bibsource    = {dblp computer science bibliography, https://dblp.org}
}

@inproceedings{Hirzle-CHI19,
  author       = {Teresa Hirzle and
                  Jan Gugenheimer and
                  Florian Geiselhart and
                  Andreas Bulling and
                  Enrico Rukzio},
  editor       = {Stephen A. Brewster and
                  Geraldine Fitzpatrick and
                  Anna L. Cox and
                  Vassilis Kostakos},
  title        = {A Design Space for Gaze Interaction on Head-mounted Displays},
  booktitle    = {Proceedings of the 2019 {CHI} Conference on Human Factors in Computing
                  Systems, {CHI} 2019, Glasgow, Scotland, UK, May 04-09, 2019},
  pages        = {625},
  publisher    = {{ACM}},
  year         = {2019},
  url          = {https://doi.org/10.1145/3290605.3300855},
  doi          = {10.1145/3290605.3300855},
  bibsource    = {dblp computer science bibliography, https://dblp.org}
}

@inproceedings{Hirzle-ETRA20,
  author       = {Teresa Hirzle and
                  Maurice Cordts and
                  Enrico Rukzio and
                  Andreas Bulling},
  editor       = {Andreas Bulling and
                  Anke Huckauf and
                  Eakta Jain and
                  Ralph Radach and
                  Daniel Weiskopf},
  title        = {A Survey of Digital Eye Strainin Gaze-Based Interactive Systems},
  booktitle    = {Proc.~ETRA},
  pages        = {9:1--9:12},
  publisher    = {{ACM}},
  year         = {2020},
  url          = {https://doi.org/10.1145/3379155.3391313},
  doi          = {10.1145/3379155.3391313},
  bibsource    = {dblp computer science bibliography, https://dblp.org}
}

@inproceedings{Feit-CHI17,
  author       = {Anna Maria Feit and
                  Shane Williams and
                  Arturo Toledo and
                  Ann Paradiso and
                  Harish Kulkarni and
                  Shaun K. Kane and
                  Meredith Ringel Morris},
  editor       = {Gloria Mark and
                  Susan R. Fussell and
                  Cliff Lampe and
                  m. c. schraefel and
                  Juan Pablo Hourcade and
                  Caroline Appert and
                  Daniel Wigdor},
  title        = {Toward Everyday Gaze Input: Accuracy and Precision of Eye Tracking
                  and Implications for Design},
  booktitle    = {Proc.~CHI},
  pages        = {1118--1130},
  publisher    = {{ACM}},
  year         = {2017},
  url          = {https://doi.org/10.1145/3025453.3025599},
  doi          = {10.1145/3025453.3025599},
  bibsource    = {dblp computer science bibliography, https://dblp.org}
}

@inproceedings{Lu-VR20,
  author       = {Feiyu Lu and
                  Shakiba Davari and
                  Lee Lisle and
                  Yuan Li and
                  Doug A. Bowman},
  title        = {Glanceable {AR:} Evaluating Information Access Methods for Head-Worn
                  Augmented Reality},
  booktitle    = {Proc.~VR},
  pages        = {930--939},
  publisher    = {{IEEE}},
  year         = {2020},
  url          = {https://doi.org/10.1109/VR46266.2020.1581100361198},
  doi          = {10.1109/VR46266.2020.1581100361198},
  bibsource    = {dblp computer science bibliography, https://dblp.org}
}

@inproceedings{Lu-VR21,
  author       = {Feiyu Lu and
                  Doug A. Bowman},
  title        = {Evaluating the Potential of Glanceable {AR} Interfaces for Authentic
                  Everyday Uses},
  booktitle    = {Proc.~VR},
  pages        = {768--777},
  publisher    = {{IEEE}},
  year         = {2021},
  url          = {https://doi.org/10.1109/VR50410.2021.00104},
  doi          = {10.1109/VR50410.2021.00104},
  bibsource    = {dblp computer science bibliography, https://dblp.org}
}

@inproceedings{Lu-CHI22,
  author       = {Feiyu Lu and
                  Yan Xu},
  editor       = {Simone D. J. Barbosa and
                  Cliff Lampe and
                  Caroline Appert and
                  David A. Shamma and
                  Steven Mark Drucker and
                  Julie R. Williamson and
                  Koji Yatani},
  title        = {Exploring Spatial {UI} Transition Mechanisms with Head-Worn Augmented
                  Reality},
  booktitle    = {Proc.~CHI},
  pages        = {550:1--550:16},
  publisher    = {{ACM}},
  year         = {2022},
  url          = {https://doi.org/10.1145/3491102.3517723},
  doi          = {10.1145/3491102.3517723},
  bibsource    = {dblp computer science bibliography, https://dblp.org}
}

@inproceedings{Renner-3DUI17,
  author       = {Patrick Renner and
                  Thies Pfeiffer},
  editor       = {Maud Marchal and
                  Robert J. Teather and
                  Bruce H. Thomas},
  title        = {Attention guiding techniques using peripheral vision and eye tracking
                  for feedback in augmented-reality-based assistance systems},
  booktitle    = {Proc.~3DUI},
  pages        = {186--194},
  publisher    = {{IEEE} Computer Society},
  year         = {2017},
  url          = {https://doi.org/10.1109/3DUI.2017.7893338},
  doi          = {10.1109/3DUI.2017.7893338},
}

@inproceedings{Gruenefeld-MobileHCI18,
  author       = {Uwe Gruenefeld and
                  Tim Claudius Stratmann and
                  Lars Pr{\"{a}}del and
                  Wilko Heuten},
  title        = {MonoculAR: a radial light display to point towards out-of-view objects
                  on augmented reality devices},
  booktitle    = {Proc.~MobileHCI Adjunct},
  pages        = {16--22},
  publisher    = {{ACM}},
  year         = {2018},
  url          = {https://doi.org/10.1145/3236112.3236115},
  doi          = {10.1145/3236112.3236115},
}

@inproceedings{Gruenefeld-ISMAR18,
  author       = {Uwe Gruenefeld and
                  Tim Claudius Stratmann and
                  Jinki Jung and
                  Hyeopwoo Lee and
                  Jeehye Choi and
                  Abhilasha Nanda and
                  Wilko Heuten},
  title        = {Guiding Smombies: Augmenting Peripheral Vision with Low-Cost Glasses
                  to Shift the Attention of Smartphone Users},
  booktitle    = {Proc.~ISMAR Adjunct},
  pages        = {127--131},
  publisher    = {{IEEE}},
  year         = {2018},
  url          = {https://doi.org/10.1109/ISMAR-Adjunct.2018.00050},
  doi          = {10.1109/ISMAR-Adjunct.2018.00050},
}

@article{Murray-Smith-Interactions22,
  author       = {Roderick Murray{-}Smith and
                  Antti Oulasvirta and
                  Andrew Howes and
                  J{\"{o}}rg M{\"{u}}ller and
                  Aleksi Ikkala and
                  Miroslav Bachinski and
                  Arthur Fleig and
                  Florian Fischer and
                  Markus Klar},
  title        = {What simulation can do for {HCI} research},
  journal      = {Interactions},
  volume       = {29},
  number       = {6},
  pages        = {48--53},
  year         = {2022},
  url          = {https://doi.org/10.1145/3564038},
  doi          = {10.1145/3564038},
}

@inproceedings{DiVerdi-ISMAR04,
  author       = {Stephen DiVerdi and
                  Tobias H{\"{o}}llerer and
                  Richard Schreyer},
  title        = {Level of Detail Interfaces},
  booktitle    = {Proc.~ISMAR},
  pages        = {300--301},
  publisher    = {{IEEE} Computer Society},
  year         = {2004},
  url          = {https://doi.org/10.1109/ISMAR.2004.38},
  doi          = {10.1109/ISMAR.2004.38},
}

@inproceedings{Hirzle-CHI21,
  author       = {Teresa Hirzle and
                  Maurice Cordts and
                  Enrico Rukzio and
                  Jan Gugenheimer and
                  Andreas Bulling},
  editor       = {Yoshifumi Kitamura and
                  Aaron Quigley and
                  Katherine Isbister and
                  Takeo Igarashi and
                  Pernille Bj{\o}rn and
                  Steven Mark Drucker},
  title        = {A Critical Assessment of the Use of {SSQ} as a Measure of General
                  Discomfort in {VR} Head-Mounted Displays},
  booktitle    = {Proc.~CHI},
  pages        = {530:1--530:14},
  publisher    = {{ACM}},
  year         = {2021},
  url          = {https://doi.org/10.1145/3411764.3445361},
  doi          = {10.1145/3411764.3445361},
}

@misc{Kirillov2023,
      title={Segment Anything}, 
      author={Alexander Kirillov and Eric Mintun and Nikhila Ravi and Hanzi Mao and Chloe Rolland and Laura Gustafson and Tete Xiao and Spencer Whitehead and Alexander C. Berg and Wan-Yen Lo and Piotr Dollár and Ross Girshick},
      year={2023},
      eprint={2304.02643},
      archivePrefix={arXiv},
      primaryClass={cs.CV}
}

@article{Grubert2017,
  author       = {Jens Grubert and
                  Tobias Langlotz and
                  Stefanie Zollmann and
                  Holger Regenbrecht},
  title        = {Towards Pervasive Augmented Reality: Context-Awareness in Augmented
                  Reality},
  journal      = {{IEEE} Trans. Vis. Comput. Graph.},
  volume       = {23},
  number       = {6},
  pages        = {1706--1724},
  year         = {2017},
  url          = {https://doi.org/10.1109/TVCG.2016.2543720},
  doi          = {10.1109/TVCG.2016.2543720},
}

@inproceedings{Lindlbauer-UIST19,
  author       = {David Lindlbauer and
                  Anna Maria Feit and
                  Otmar Hilliges},
  title        = {Context-Aware Online Adaptation of Mixed Reality Interfaces},
  booktitle    = {Proc.~UIST},
  pages        = {147--160},
  publisher    = {{ACM}},
  year         = {2019},
  url          = {https://doi.org/10.1145/3332165.3347945},
  doi          = {10.1145/3332165.3347945}
}

@article{Arora-2025,
  author       = {Parth Arora and
                  Ethan I Kimmel and
                  Katherine Huang and
                  Tyler Kwok and
                  Yukun Song and
                  Sofia Anandi Vempala and
                  Georgianna Lin and
                  Ozan Cakmakci and
                  Thad Starner},
  title        = {Positioning Monocular Optical See Through Head Worn Displays in Glasses
                  for Everyday Wear},
  journal      = {CoRR},
  volume       = {abs/2505.09047},
  year         = {2025},
  url          = {https://doi.org/10.48550/arXiv.2505.09047},
  doi          = {10.48550/ARXIV.2505.09047},
  eprinttype    = {arXiv},
  eprint       = {2505.09047},
}

@article{HaynesS-IMWUT17,
  author       = {Malcolm Gibran Haynes and
                  Thad Starner},
  title        = {Effects of Lateral Eye Displacement on Comfort While Reading from
                  a Video Display Terminal},
  journal      = {Proc. {ACM} Interact. Mob. Wearable Ubiquitous Technol.},
  volume       = {1},
  number       = {4},
  pages        = {138:1--138:17},
  year         = {2017},
  url          = {https://doi.org/10.1145/3161177},
  doi          = {10.1145/3161177},
}

@inproceedings{chen2023imperceptible,
  author    = {Chen, Kaizhan and Duinkharjav, Batchuluun and Ujjainkar, Nilay and Shahan, Ensar and Tyagi, Akanksha and He, Jing and Sun, Qian},
  title     = {Imperceptible Color Modulation for Power Saving in VR/AR},
  booktitle = {ACM SIGGRAPH 2023 Emerging Technologies},
  year      = {2023},
  pages     = {1--2}
}

@article{duinkharjav2022color,
  author  = {Duinkharjav, Batchuluun and Chen, Kaizhan and Tyagi, Akanksha and He, Jing and Zhu, Yixin and Sun, Qian},
  title   = {Color-Perception-Guided Display Power Reduction for Virtual Reality},
  journal = {ACM Transactions on Graphics (TOG)},
  volume  = {41},
  number  = {6},
  year    = {2022},
  pages   = {1--16}
}

@inproceedings{koulieris2019near,
  author    = {Koulieris, Georgios A. and Akşit, Kemal and Stengel, Michael and Mantiuk, Rafal K. and Mania, Katarzyna and Richardt, Christian},
  title     = {Near-Eye Display and Tracking Technologies for Virtual and Augmented Reality},
  booktitle = {Computer Graphics Forum},
  year      = {2019},
  volume    = {38},
  number    = {2},
  pages     = {493--519}
}

@inproceedings{ku2019peritext,
  author    = {Ku, Pei-Shan and Lin, Yu-Chuan and Peng, Yu-Hao and Chen, Ming-Yu},
  title     = {PeriText: Utilizing Peripheral Vision for Reading Text on Augmented Reality Smart Glasses},
  booktitle = {2019 IEEE Conference on Virtual Reality and 3D User Interfaces (VR)},
  year      = {2019},
  month     = mar,
  pages     = {630--635},
  publisher = {IEEE}
}

@inproceedings{luyten2016hidden,
  author    = {Luyten, Kris and Degraen, Dries and Rovelo Ruiz, Gustavo and Coppers, Sven and Vanacken, David},
  title     = {Hidden in Plain Sight: An Exploration of a Visual Language for Near-Eye Out-of-Focus Displays in the Peripheral View},
  booktitle = {Proceedings of the 2016 CHI Conference on Human Factors in Computing Systems},
  year      = {2016},
  month     = may,
  pages     = {487--497}
}

@inproceedings{parmar2023impact,
  author    = {Parmar, Mihir and Silpasuwanchai, Chai},
  title     = {Impact of User Mobility on Attentional Tunneling in Handheld AR},
  booktitle = {Extended Abstracts of the 2023 CHI Conference on Human Factors in Computing Systems},
  year      = {2023},
  month     = apr,
  pages     = {1--6}
}

@misc{rauschnabel2015ar,
  author       = {Rauschnabel, Philipp A. and Brem, Alexander and Ro, Young},
  title        = {Augmented Reality Smart Glasses: Definition, Conceptual Insights, and Managerial Importance},
  year         = {2015},
  note         = {Working paper, College of Business, University of Michigan–Dearborn},
  howpublished = {Unpublished manuscript}
}

@inproceedings{sun2018investigating,
  author    = {Sun, Xinran and Varshney, Amitabh},
  title     = {Investigating Perception Time in the Far Peripheral Vision for Virtual and Augmented Reality},
  booktitle = {SAP},
  year      = {2018},
  month     = aug,
  pages     = {13--1}
}

@inproceedings{syiem2021impact,
  author    = {Syiem, Bryan V. and Kelly, Rebecca M. and Goncalves, Jose and Velloso, Eduardo and Dingler, Tobias},
  title     = {Impact of Task on Attentional Tunneling in Handheld Augmented Reality},
  booktitle = {Proceedings of the 2021 CHI Conference on Human Factors in Computing Systems},
  year      = {2021},
  month     = may,
  pages     = {1--14}
}

@inproceedings{tanuwidjaja2014chroma,
  author    = {Tanuwidjaja, Erin and Huynh, Duong and Koa, Khoa and Nguyen, Cuong and Shao, Cheng and Torbett, Patrick and Weibel, Nico},
  title     = {Chroma: A Wearable Augmented-Reality Solution for Color Blindness},
  booktitle = {Proceedings of the 2014 ACM International Joint Conference on Pervasive and Ubiquitous Computing},
  year      = {2014},
  month     = sep,
  pages     = {799--810}
}

@inproceedings{trepkowski2021multisensory,
  author    = {Trepkowski, Christoph and Marquardt, Andreas and Eibich, Thies D. and Shikanai, Yusuke and Maiero, Jens and Kiyokawa, Kiyoshi and König, Peter},
  title     = {Multisensory Proximity and Transition Cues for Improving Target Awareness in Narrow Field of View Augmented Reality Displays},
  booktitle = {IEEE Transactions on Visualization and Computer Graphics},
  volume    = {28},
  number    = {2},
  year      = {2021},
  pages     = {1342--1362}
}

@article{strasburger2011peripheral,
  author  = {Strasburger, Hans and Rentschler, Ingo and J\"uttner, Martin},
  title   = {Peripheral Vision and Pattern Recognition: A Review},
  journal = {Journal of Vision},
  volume  = {11},
  number  = {5},
  pages   = {13--13},
  year    = {2011}
}

@inproceedings{davari2022validating,
  author    = {Davari, Seyed and Lu, Feiyu and Bowman, Doug A.},
  title     = {Validating the Benefits of Glanceable and Context-Aware Augmented Reality for Everyday Information Access Tasks},
  booktitle = {2022 IEEE Conference on Virtual Reality and 3D User Interfaces (VR)},
  year      = {2022},
  month     = mar,
  pages     = {436--444},
  publisher = {IEEE}
}

@inproceedings{janaka2022paracentral,
  author    = {Nuwan Janaka and Paul Strohmeier and Dongwook Yoon and Yutaka Tokuda and Wei Sun and Kai Kunze},
  title     = {Paracentral and Near-Peripheral Visualizations: Towards Attention-Maintaining Secondary Information Presentation on OHMDs during In-Person Social Interactions},
  booktitle = {Proceedings of the 2022 CHI Conference on Human Factors in Computing Systems},
  year      = {2022},
  publisher = {Association for Computing Machinery},
  address   = {New York, NY, USA},
  pages     = {1--18},
  doi       = {10.1145/3491102.3517681}
}

@book{bowman2021,
  author    = {Doug A. Bowman and Ernst Kruijff and Joseph J. LaViola Jr. and Ivan Poupyrev},
  title     = {3D User Interfaces: Theory and Practice},
  edition   = {2nd},
  year      = {2021},
  publisher = {Addison-Wesley},
  address   = {Boston, MA, USA},
  isbn      = {978-0-13-464346-2}
}

@article{pai2019handsfree,
  author    = {Yun Suen Pai and Tilman Dingler and Kai Kunze},
  title     = {Assessing Hands-Free Interactions for VR Using Eye Gaze and Electromyography},
  journal   = {Virtual Reality},
  volume    = {23},
  number    = {2},
  pages     = {119--131},
  year      = {2019},
  publisher = {Springer},
  doi       = {10.1007/s10055-018-0341-3}
}

@inproceedings{pfeuffer2017gazepinch,
  author    = {Ken Pfeuffer and Benedikt Mayer and Diako Mardanbegi and Hans Gellersen},
  title     = {Gaze+ Pinch Interaction in Virtual Reality},
  booktitle = {Proceedings of the 5th Symposium on Spatial User Interaction (SUI '17)},
  year      = {2017},
  pages     = {99--108},
  publisher = {Association for Computing Machinery},
  address   = {New York, NY, USA},
  doi       = {10.1145/3131277.3132180}
}

@inproceedings{sidenmark2020outline,
  author    = {Ludwig Sidenmark and Christopher Clarke and Xuesong Zhang and Jenny Phu and Hans Gellersen},
  title     = {Outline Pursuits: Gaze-Assisted Selection of Occluded Objects in Virtual Reality},
  booktitle = {Proceedings of the 2020 CHI Conference on Human Factors in Computing Systems (CHI '20)},
  year      = {2020},
  pages     = {1--13},
  publisher = {Association for Computing Machinery},
  address   = {New York, NY, USA},
  doi       = {10.1145/3313831.3376873}
}

@inproceedings{sidenmark2019eyehead,
  author    = {Ludwig Sidenmark and Hans Gellersen},
  title     = {Eye\&Head: Synergetic Eye and Head Movement for Gaze Pointing and Selection},
  booktitle = {Proceedings of the 32nd Annual ACM Symposium on User Interface Software and Technology (UIST '19)},
  year      = {2019},
  pages     = {1161--1174},
  publisher = {Association for Computing Machinery},
  address   = {New York, NY, USA},
  doi       = {10.1145/3332165.3347945}
}

@inproceedings{chen2023gazeraycursor,
  author    = {Di Laura Chen and Yuzhen Chen and Michelle Annett and Daniel Vogel},
  title     = {GazeRayCursor: Facilitating Virtual Reality Target Selection by Blending Gaze and Controller Raycasting},
  booktitle = {Proceedings of the 29th ACM Symposium on Virtual Reality Software and Technology (VRST '23)},
  year      = {2023},
  pages     = {1--11},
  publisher = {Association for Computing Machinery},
  address   = {New York, NY, USA},
  doi       = {10.1145/3611659.3615696}
}

@article{nothdurft1993preattentive,
  author  = {Nothdurft, Hans-Christoph},
  title   = {The role of features in preattentive vision: Comparison of orientation, motion and color cues},
  journal = {Vision Research},
  year    = {1993},
  volume  = {33},
  number  = {14},
  pages   = {1937--1958},
  doi     = {10.1016/0042-6989(93)90020-W}
}

@article{andersen2019attentional,
  author  = {Andersen, Emil and Maier, Anja},
  title   = {The attentional guidance of individual colours in increasingly complex displays},
  journal = {Applied Ergonomics},
  year    = {2019},
  volume  = {81},
  pages   = {102885},
  doi     = {10.1016/j.apergo.2019.102885}
}

@incollection{wolfe2018visual,
  author    = {Wolfe, Jeremy M.},
  title     = {Visual search},
  booktitle = {Stevens' Handbook of Experimental Psychology and Cognitive Neuroscience},
  editor    = {Wixted, John T.},
  publisher = {John Wiley \& Sons},
  year      = {2018},
  volume    = {2},
  chapter   = {13},
  pages     = {1--55},
  doi       = {10.1002/9781119170174.epcn213}
}

@article{clifford2010color,
  title        = {Color categories affect pre-attentive color perception},
  author       = {Clifford, Alexandra and Spehar, Branka and Franklin, Anna},
  journal      = {Biological Psychology},
  volume       = {85},
  number       = {2},
  pages        = {275--282},
  year         = {2010},
  publisher    = {Elsevier},
  doi          = {10.1016/j.biopsycho.2010.07.009}
}

@article{khomeriki2025attention,
  title        = {Visual Attention Distribution According to Size, Color, and Spatial Location of Stimuli under Foveal and Peripheral Vision Conditions},
  author       = {Khomeriki, Manana and Lomashvili, Natela},
  journal      = {ESI Preprints (European Scientific Journal, ESJ)},
  volume       = {21},
  number       = {15},
  pages        = {22--22},
  year         = {2025}
}

@article{sharma2005ciede2000,
  title        = {The {CIEDE2000} color-difference formula: Implementation notes, supplementary test data, and mathematical observations},
  author       = {Sharma, Gaurav and Wu, Wencheng and Dalal, Edul N.},
  journal      = {Color Research \& Application},
  volume       = {30},
  number       = {1},
  pages        = {21--30},
  year         = {2005},
  publisher    = {Wiley},
}

@article{khanam2024yolov11,
  title   = {Yolov11: An Overview of the Key Architectural Enhancements},
  author  = {Khanam, Rahima and Hussain, Muhammad},
  journal = {arXiv preprint arXiv:2410.17725},
  year    = {2024}
}

@inproceedings{Salvucci-ETRA00,
  author       = {Dario D. Salvucci and
                  Joseph H. Goldberg},
  editor       = {Andrew T. Duchowski},
  title        = {Identifying fixations and saccades in eye-tracking protocols},
  booktitle    = {Proceedings of the Eye Tracking Research {\&} Application Symposium,
                  {ETRA} 2000, Palm Beach Gardens, Florida, USA, November 6-8, 2000},
  pages        = {71--78},
  publisher    = {{ACM}},
  year         = {2000},
  url          = {https://doi.org/10.1145/355017.355028},
}

@inproceedings{Vermeulen-CHI13,
  author    = {Jo Vermeulen and Kris Luyten and Elise van den Hoven and Karin Coninx},
  title     = {Crossing the Bridge over Norman's Gulf of Execution: Revealing Feedforward's True Identity},
  booktitle = {Proceedings of the SIGCHI Conference on Human Factors in Computing Systems (CHI '13)},
  pages     = {1931--1940},
  year      = {2013},
  publisher = {ACM},
  address   = {New York, NY, USA},
  isbn      = {978-1-4503-1899-0},
  doi       = {10.1145/2470654.2466255}
}

@inproceedings{Guillon-CHI15,
  author    = {Maxime Guillon and Fran{\c c}ois Leitner and Laurence Nigay},
  title     = {Investigating Visual Feedforward for Target Expansion Techniques},
  booktitle = {Proceedings of the 33rd Annual ACM Conference on Human Factors in Computing Systems (CHI '15)},
  pages     = {2777--2786},
  year      = {2015},
  publisher = {ACM},
  address   = {New York, NY, USA},
  isbn      = {978-1-4503-3145-6},
  doi       = {10.1145/2702123.2702375}
}

@inproceedings{Sadasivan-CHI05,
  author    = {Sajay Sadasivan and Randall L. Morrison and Michael J. Dorneich and J. Gregory Trafton},
  title     = {Use of Eye Movements as Feedforward Training for a Synthetic Aircraft Inspection Task},
  booktitle = {Proceedings of the SIGCHI Conference on Human Factors in Computing Systems (CHI '05)},
  pages     = {141--150},
  year      = {2005},
  publisher = {ACM},
  address   = {New York, NY, USA},
  isbn      = {1-58113-998-5},
  doi       = {10.1145/1054972.1054993}
}

@inproceedings{Yu-CHI24,
  author    = {Xingyao Yu and Benjamin Lee and Michael Sedlmair},
  title     = {Design Space of Visual Feedforward and Corrective Feedback in {XR}-Based Motion Guidance Systems},
  booktitle = {Proceedings of the 2024 {CHI} Conference on Human Factors in Computing Systems (CHI '24)},
  pages     = {723:1--723:15},
  year      = {2024},
  publisher = {ACM},
  address   = {New York, NY, USA},
  isbn      = {979-8-4007-0330-0},
  doi       = {10.1145/3613904.3642143}
}

@article{Nikolic-HF01,
  author    = {Mark I. Nikolic and Nadine B. Sarter},
  title     = {Peripheral Visual Feedback: A Powerful Means of Supporting Effective Attention Allocation in Event-Driven, Data-Rich Environments},
  journal   = {Human Factors},
  volume    = {43},
  number    = {1},
  pages     = {30--38},
  year      = {2001},
  publisher = {SAGE Publications},
  doi       = {10.1518/001872001775992525}
}

@article{Wang-JCISE24,
  author    = {I{-}Jan Wang and Hao{-}Cheng Kuo and Cheng{-}Kai Huang and Yi{-}Liang Chuang and Ting{-}Han Fan and Yu{-}Ting Liu and Hao{-}Ting Chen and Liwei Chan},
  title     = {Assistive Sensory Feedback for Trajectory Tracking in Augmented Reality},
  journal   = {Journal of Computing and Information Science in Engineering},
  volume    = {24},
  number    = {3},
  pages     = {031001},
  year      = {2024},
  publisher = {ASME},
  doi       = {10.1115/1.4065218}
}

@inproceedings{Richards-SUI19,
  author    = {Kendra Richards and Danilo Gasques and Jay Peters and Bruce H. Thomas},
  title     = {Analysis of Peripheral Vision and Vibrotactile Feedback During Proximal Search Tasks with Dynamic Virtual Entities in Augmented Reality},
  booktitle = {Proceedings of the Symposium on Spatial User Interaction (SUI '19)},
  pages     = {1--9},
  year      = {2019},
  publisher = {ACM},
  address   = {New York, NY, USA},
  isbn      = {978-1-4503-6893-4},
  doi       = {10.1145/3357251.3357581}
}

@article{Fernandes-IJHCI25,
  author    = {Ajoy S. Fernandes and Ken Pfeuffer and I. Scott MacKenzie},
  title     = {Gaze Inputs for Targeting: The Eyes Have It, Not With a Cursor},
  journal   = {International Journal of Human--Computer Interaction},
  pages     = {1--19},
  year      = {2025},
  publisher = {Taylor \& Francis},
  doi       = {10.1080/10447318.2024.2419477}
}

\clearpage
\appendix

\section{Color Enhancement Algorithm}
\label{app:algorithm}

Our color enhancement algorithm entails three steps: \textit{(1)} quantization, \textit{(2)} palette-level color distance analysis, and \textit{(3)} masked color enhancement. Fig.~\ref{fig:color enhancement pipeline}

\p{Quantization}
In the first stage, both target and reference textures are reduced to a compact set of dominant colors. Using MiniBatchKMeans with $K=7$, each pixel is reassigned to its nearest centroid, yielding a quantized image and a histogram of cluster coverages. This step highlights perceptually salient palettes consistent with peripheral vision's reduced sensitivity to detail, while also lowering computational load.

\p{Palette-Level Color Distance Analysis}
Next, we compare the palettes of the target and reference textures to determine the extent of enhancement needed for the target. We compute weighted perceptual distances across all pairs of palette centroids:
\[
\Delta E_{\mathrm{total}} = \sum_{i=1}^{7}\sum_{j=1}^{7} h_T(i)\,h_R(j)\,\Delta E_{00}\!\bigl(c_i^{T},c_j^{R}\bigr),
\]
where $c_i^{T}$ and $c_j^{R}$ are the $i$-th and $j$-th centroids of the target and reference palettes, $h_T(i)$ and $h_R(j)$ are their normalized cluster weights, and $\Delta E_{00}(c_i^{T},c_j^{R})$ is the CIEDE2000 color difference~\cite{sharma2005ciede2000} between the two centroids in CIE~Lab space. We use CIEDE2000 because it captures perceptual non-uniformities, providing differences that better reflect how humans perceive changes in lightness, chroma, and hue compared to simpler metrics such as Euclidean distance. Analogous computations yield $\Delta L$ for lightness and $\Delta C$ for chroma differences. Together, these metrics quantify how distinct the two textures are in overall appearance.

For enhancement preparation, we also identify a \emph{shared dominant color}, chosen from the reference palette but also prominent in the target. This color acts as an anchor, marking the overlap where confusion is most likely in peripheral vision. By centering enhancement around this shared color, the algorithm selectively increases contrast where target and reference would otherwise appear too similar. The global distance metrics ($\Delta E_{\mathrm{total}}, \Delta L, \Delta C$) together with the shared dominant color form the basis for constructing the boost map in the next stage.

\p{Masked Color Enhancement}
In the final stage, enhancement is applied selectively to regions of the target associated with the shared dominant color. A similarity mask is computed by measuring the CIEDE2000 distance from each target pixel to this color. Scaled by the global perceptual distance $\Delta E_{\mathrm{total}}$, the mask produces a \emph{boost map} $B(x,y)$ that localizes enhancement to areas where target and reference overlap most strongly. This ensures adjustments increase perceptual contrast precisely where peripheral confusion is most likely. While $\Delta E_{\mathrm{total}}$ controls overall enhancement strength, the relative contributions of $\Delta L$ and $\Delta C$ determine whether emphasis falls on lightness or chroma.

We capture this weighting by
\[
\alpha = \frac{\Delta C}{\Delta L + \Delta C},
\]
where $\alpha$ biases enhancement toward chroma when $\Delta C$ dominates and toward luminance when $\Delta L$ dominates. Saturation is applied as a general vividness boost, independent of $\alpha$. With this weighting, the three channels are adjusted as follows:

\begin{itemize}
  \item \textbf{Luminance:} brightness is lifted by scaling Lab lightness:
  \[
  L'(x,y) = L(x,y)\,\bigl[1 + B(x,y)(\text{max\_luminance}-1)(1-\alpha)\bigr],
  \]
  where \texttt{max\_luminance} caps the maximum increase. Gamma correction and CLAHE (Contrast Limited Adaptive Histogram Equalization) preserve highlight detail and local contrast.

  \item \textbf{Saturation:} HSV saturation is boosted by
  \[
  S'(x,y) = S(x,y)\,\bigl[1 + B(x,y)(\text{max\_sat\_boost}-1)\alpha\bigr],
  \]
  with \texttt{max\_sat\_boost} controlling the maximum gain. Saturation is weighted by $\alpha$ so that it contributes more when chromatic differences ($\Delta C$) dominate, and less when luminance differences ($\Delta L$) are the primary factor. This ensures that saturation lifting is emphasized only when it is most perceptually useful.

  \item \textbf{Chroma push:} Lab chroma values $(a,b)$ are shifted outward from the reference mean $\mu_{ab}$:
  \[
  (a',b')(x,y) = (a,b)(x,y) + B(x,y)\cdot \text{ab\_push}\cdot \frac{(a,b)(x,y) - \mu_{ab}}{\|(a,b)(x,y) - \mu_{ab}\|_2},
  \]
  where \texttt{ab\_push} sets the maximum directional shift. Unlike saturation, chroma push operates independently of $\alpha$, because its role is to explicitly separate the target's hue from the reference. This guarantees local color contrast even when overall luminance differences are dominant.
\end{itemize}

The result is a color-enhanced target image that is brighter, more vivid, and chromatically distinct from the reference image, allowing participants to associate the peripheral proxy with the correct real-world target at a glance.

\end{document}